\documentclass[aps,prl, twocolumn,floatfix,superscriptaddress,preprintnumbers,
               tightenlines,showpacs,showkeys,nofootinbib,notoccite]{revtex4-1}
\usepackage[utf8]{inputenc}
\usepackage[colorlinks=true,citecolor=blue,linkcolor=blue]{hyperref}
\usepackage[normalem]{ulem}
\usepackage{amsmath,amssymb, mathrsfs}
\usepackage{epsfig}
\usepackage{graphicx}               
\usepackage{url}
\usepackage{color}
\usepackage{slashed}
\usepackage{multirow}
\usepackage{placeins}
\usepackage[dvipsnames]{xcolor}
\usepackage{epstopdf}
\usepackage{soul}
\usepackage{tikz}
\usepackage[capitalise, english]{cleveref}
\usepackage{siunitx}
\usepackage{xspace}
\usepackage{booktabs}
\usetikzlibrary{trees}
\usetikzlibrary{decorations.pathmorphing}
\usetikzlibrary{decorations.markings}
\usepackage{orcidlink}
\usepackage{array}
\usepackage{tabularx}
\usepackage{amssymb}
\usepackage{amsfonts}
\newcommand\myshade{80}
\colorlet{mylinkcolor}{ForestGreen}
\colorlet{mycitecolor}{Red}
\colorlet{myurlcolor}{violet}

\usepackage{bbold}

\hypersetup{
  linkcolor  = mylinkcolor!\myshade!black,
  citecolor  = mycitecolor!\myshade!black,
  urlcolor   = myurlcolor!\myshade!black,
  colorlinks = true
}
\def\mET{E_T \hspace{-1.0em}/\;\:}
\definecolor{jblue}{RGB}{20,50,100}
\definecolor{npurple}{RGB} {153, 51, 204}
\definecolor{wred}{RGB}{217,0,56}
\definecolor{white}{RGB}{255,255,255}

\definecolor{korange}{RGB}{235, 80,  43}
\definecolor{korange2}{RGB}{245, 100,  63}
\definecolor{kyelloworange}{RGB}{255, 210,  110}
\definecolor{kyelloworange2}{RGB}{240, 170,  90}
\definecolor{kred}{RGB}{204,  102, 153}
\definecolor{kpurple}{RGB}{153,  61, 190}
\definecolor{kpurplelight}{RGB}{213,  161, 230}

 \definecolor{tobycolour}{rgb}{.5,.0,.5}

\DeclareSIUnit\year{yr}
\DeclareSIUnit\pc{pc}
\DeclareSIUnit\ergs{ergs}
\DeclareSIUnit\msun{\ensuremath{M_\odot}}
\allowdisplaybreaks

\renewcommand{\vec}[1]{{\mathbf{#1}}}

\makeatletter
\providecommand*{\diff}%
  {\@ifnextchar^{\DIfF}{\DIfF^{}}}
\def\DIfF^#1{%
  \mathop{\mathrm{\mathstrut d}}%
    \nolimits^{#1}\gobblespace}
\def\gobblespace{%
  \futurelet\diffarg\opspace}
\def\opspace{%
  \let\DiffSpace\!%
  \ifx\diffarg(%
    \let\DiffSpace\relax
  \else
    \ifx\diffarg[%
      \let\DiffSpace\relax
    \else
        \ifx\diffarg\{%
        \let\DiffSpace\relax
      \fi\fi\fi\DiffSpace}

\usepackage{tikz,xcolor,hyperref}

\definecolor{lime}{HTML}{A6CE39}
\DeclareRobustCommand{\orcidicon}{\hspace{-1mm}
	\begin{tikzpicture}
	\draw[lime, fill=lime] (0,0) 
	circle [radius=0.16] 
	node[white] {{\fontfamily{qag}\selectfont \tiny \,ID}};
	\draw[white, fill=white] (-0.0525,0.095) 
	circle [radius=0.007];
	\end{tikzpicture}
	\hspace{-3mm}
}

\foreach \x in {A, ..., Z}{\expandafter\xdef\csname orcid\x\endcsname{\noexpand\href{https://orcid.org/\csname orcidauthor\x\endcsname}
			{\noexpand\orcidicon}}
}

\usepackage{xcolor}      


\begin{document}
\preprint{}

\title{One Event, Two Frontiers: From LZ to LHC}

\author{Md Sadique Anwar}
\email{sadique.anwar@saha.ac.in}
\affiliation{Saha Institute of Nuclear Physics, 
1/AF Bidhannagar, Kolkata 700064, India} 
\affiliation{Homi Bhabha National Institute, Training School Complex,
Anushakti Nagar, Mumbai 400094, India}
\author{Nivedita Ghosh\orcidlink{0000-0001-8469-3227}} 
\email{nivedita.ghosh@ipmu.jp}
\affiliation{Kavli IPMU (WPI), UTIAS, University of Tokyo,  Kashiwa, Chiba, 277-8583, Japan}
\author{Sahabub Jahedi\orcidlink{0000-0003-1016-8264}} 
\email{sahabub@m.scnu.edu.cn}
\affiliation{State Key Laboratory of Nuclear Physics and
Technology, Institute of Quantum Matter, South China Normal
University, Guangzhou 510006, China}
\affiliation{Guangdong Basic Research Center of Excellence for
Structure and Fundamental Interactions of Matter, Guangdong
Provincial Key Laboratory of Nuclear Science, Guangzhou
510006, China}
\author{Tarak Nath Maity\orcidlink{0000-0003-1069-6358}}
\email{tarak.maity.physics@gmail.com}
\affiliation{Saha Institute of Nuclear Physics, 1/AF Bidhannagar, Kolkata 700064, India} 
\affiliation{Homi Bhabha National Institute, Training School Complex, Anushakti Nagar, Mumbai 400094, India}
\date{\today}


\begin{abstract}
The 248 keV nuclear-recoil event observed by the LUX-ZEPLIN (LZ) experiment could potentially arise from dark matter (DM) interactions with nucleons. We investigate elastic scattering interpretation, focusing on Lagrangians that yield a local significance of $3\sigma$ or higher for some DM mass. We identify DM parameter space compatible with the LZ event and assess its consistency with collider constraints by recasting the ATLAS monojet plus missing-transverse-momentum search at $\sqrt{s}=13$ TeV, based on $139~\mathrm{fb}^{-1}$ of data. For most of the effective interactions, we find that LHC constraints exclude the DM interpretation of the LZ event for DM masses below $\lesssim\,$TeV subject to the effective field theory (EFT) validity. In particular, we derive, for the first time, LHC constraints on several explicit momentum-dependent Lagrangians, including $\mathcal{L}_{10}$. For many interactions, these constraints are sufficiently strong that the corresponding parameter space below DM mass $\lesssim\,$TeV may be challenging to probe even with future direct detection experiments. Our results, therefore, have broader implications beyond the interpretation of the LZ event itself.
\end{abstract}

\maketitle

\setcounter{footnote}{0}  
\renewcommand{\thefootnote}{\arabic{footnote}} 

\section{Introduction}
\label{sec:introduction}  
The first-generation DM direct detection experiments typically searched for the elastic scattering of DM with nucleons \cite{LZ:2024zvo, XENON:2025vwd, PandaX:2024qfu}. One of the leading experiments in this pursuit, LZ, has reported an event with a recoil energy of $248 \pm 23\,({\rm stat}) \pm 23\,({\rm syst})$ keV at around $1.5\sigma$ away from the nuclear-recoil band\,\cite{LZ:2026axp}. This event is difficult to accommodate within the expected background, naturally raising the possibility of DM interpretation. If interpreted as a DM signal, this could potentially indicate the presence of a TeV-scale particle DM with interactions with nucleons.

Several types of DM–nucleon interactions could potentially explain such a signal. Two leading possibilities are, first, intrinsically inelastic scattering between states in the dark sector with nucleon\,\cite{LZ:2026axp, Fan:2026kxx, Freese:2026sga, Wu:2026nhi, DiMauro:2026ldr, Du:2026guj, Rodd:2026tyn, Dent:2026bji, Visinelli:2026kgt, Gu:2026vto, Smirnov:2026aqk, McCabe:2026crm, Chattopadhyay:2026ryw, deLima:2026shq, Baer:2026fpy, Bose:2026ndd, Lee:2026wof, Wang:2026ytg, Yang:2026wpb, Das:2026uyy, Barman:2026omh, Ahmed:2026qjg, DiMauro:2026dqp, Du:2026lpa, Bandyopadhyay:2026gjw, Borah:2026zwf, Bisal:2026khf, Cheung:2026byg, Yuan:2026djt, Zhu:2026dag, Asadi:2026iot, Lee:2026xxh, Lee:2026jxl, Langhoff:2026ujr, Chatterjee:2026scv, Fan:2026hzw, Qi:2026vyp, Frolovsky:2026tvq, Kumar:2026lgi, Nguyen:2026lui, Sannino:2026hkc, Maier:2026qpr, Girmohanta:2026qzo, Tang:2026lgu, Nomura:2026qyq, Yuan:2026djt, Cheung:2026byg, Lee:2026jxl, Asadi:2026iot, Ahmed:2026kan, Gemmell:2026yaw, Delepine:2026ith, Langhoff:2026qgo, Murayama:2026apt, Das:2026jsq, Roy:2026wzb}, particularly motivated by the relatively high observed recoil energy, and second, elastic scattering between DM and nuclei\,\cite{LZ:2026axp, Unwin:2026rdp, Lou:2026idn, Su:2026rwz, Yamashita:2026ump, Elahi:2026vlm, Khan:2026nwp, He:2026hqz, Heikinheimo:2026kwp, Chauhan:2026udz, Jung:2026otm, Khan:2026osp, Capolupo:2026uyi, Roy:2026wzb} (see \cite{Lou:2026idn, Liang:2026coz, Okada:2026eol, Kannike:2026qyl, Chattaraj:2026fxn, Aghaie:2026vsu, Brdar:2026ukx, Kalita:2026hvb, Chanda:2026wdl} for other considerations). In the case of inelastic scattering, the additional kinematic parameters associated with the dark-sector mass splitting provide extra freedom, beyond the DM mass and coupling, to accommodate the observed signal. For elastic scattering, on the other hand, momentum- and spin-dependent interactions can provide a better fit to the signal than the conventional spin-independent (SI) interaction.

\begin{figure}[t]
     \centering
     \includegraphics[scale=0.55]{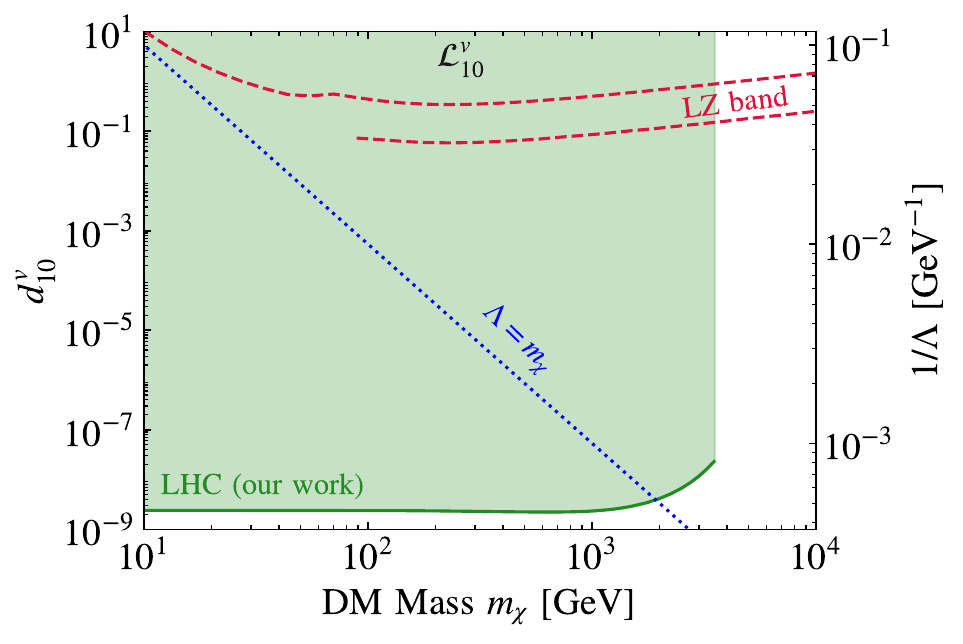}
     \caption{DM parameter space of interest for a representative Lagrangian, $L_{10}^v$ in the $m_{\chi}-d$ (or inverse of effective scale $1/\Lambda$). The red dashed band denotes the region preferred by the LZ event. The region above the green solid line is ruled out at $95\%$ confidence level by our collider analysis of monojet+$\mET$ searches using 13 TeV LHC data with an integrated luminosity of $\mathfrak{L}_{\rm int}=139~\mathrm{fb}^{-1}$. The blue dotted line corresponds to $m_{\chi}=\Lambda$ and serves as an indicative boundary for the validity of the EFT, with the EFT expected to remain valid below this line.}
     \label{fig:L_10_v}
\end{figure}

Following this excitement, several phenomenological studies have explored various aspects of inelastic DM interpretations of the LZ event. Here, instead, we take a different approach and investigate elastic DM--nucleon interactions. Depending on the spin of the DM particle, a variety of Lorentz-invariant elastic interactions between DM and nucleons can be constructed from bilinear combinations of the corresponding fields. For Dirac fermionic DM, which is the focus of this work, these interactions can be formed from the independent scalar, pseudoscalar, vector, axial-vector, and tensor bilinears of the DM and nucleon fields, subject to the appropriate Lorentz symmetry properties. These interactions give rise to a set of effective Lagrangians, several of which yield a local significance of $\geq 3\sigma$ for DM mass around TeV. We restrict our analysis to these interactions as summarized in Tab.\,\ref{tab:lz_lhc}.

The preference for TeV-scale DM is particularly interesting in light of the energy reach of the Large Hadron Collider (LHC). This motivates us to investigate whether the regions of DM parameter space that are compatible with the LZ event are also consistent with existing LHC constraints. The DM–nucleon effective interactions considered in our analysis can be consistently matched onto the DM effective field theory (DMEFT) framework\,\cite{Bishara:2016hek, Brod:2017bsw, Bhattacharya:2021edh, Liang:2023yta}, which describes the corresponding effective interactions between DM and quarks. These interactions give rise to DM production at the LHC, leading to the characteristic monojet+$\mET$ signature. Such signatures have been searched by the ATLAS Collaboration at the LHC for 13 TeV with an integrated luminosity ($\mathfrak{L}_{\rm int}$) of 139 $\rm fb^{-1}$\,\cite{ATLAS:2021kxv} and also have been extensively studied as probes of DM interaction within EFT frameworks~\cite{Racco_2015,De_Simone_2016,Bauer:2016pug,Bruggisser:2016nzw,Pobbe:2017wrj,Belwal:2017nkw,Belyaev:2018pqr,Roy:2024ear,Roy:2025pht}. CMS has also performed similar searches~\cite{CMS:2021far} for 13 TeV but with $\mathfrak{L}_{\rm int}=101~\rm fb^{-1}$. For our present work, we recast the ATLAS results of Ref.\,\cite{ATLAS:2021kxv} to constrain the DM–nucleon effective interactions considered in our analysis and thereby assess their compatibility with the parameter space favored by the LZ data.

We find that a substantial fraction of the parameter space capable of explaining the LZ event, including DM masses extending up to a few TeV, is strongly disfavored by current LHC data for most of the considered Lagrangian. For a representative interaction of the iso-vector $\mathcal{L}_{10}^v$, we show this comparison in Fig.\,\ref{fig:L_10_v}. The LZ band denotes the region of DM parameter space consistent with the observation of one event, while the light green shaded region indicates the parameter space excluded by the current LHC constraints. Consequently, current LHC data favor DM masses above a few TeV, disfavoring elastic-scattering interpretations of the LZ event below a few TeV DM with the condition that description of EFT is valid. We obtain similar results for most of the other effective Lagrangians and summarized in Tab.\,\ref{tab:lz_lhc}.

Given that the interactions favored by the LZ event are predominantly momentum- and spin-dependent, the strong LHC constraints are qualitatively consistent with expectations from previous studies of collider probes of such interactions \cite{Roy:2024ear}. However, we emphasize that, in the context of the recently reported LZ event, this complementarity between direct detection and collider constraints has not been demonstrated previously, to the best of our knowledge. Furthermore, we present the first study of current LHC constraints on explicitly momentum-dependent Lagrangians, such as $\mathcal{L}_{10}$. For DM masses below a few TeV, these constraints are orders of magnitude stronger than existing direct-detection bounds, and the corresponding parameter space may be extremely challenging to probe even in future direct-detection experiments. Our results, therefore, have broader implications beyond the DM interpretation of the LZ event. The rest of the paper is organized as follows. Section~\ref{sec:LZ} is dedicated to the analysis of the LZ event, while Section~\ref{sec:coll_ana} presents our analysis of the LHC searches. We conclude in Section~\ref{sec:conclusion}.

\section{LZ event}~\label{sec:LZ}

The LZ detector has observed a single event corresponding to a nuclear recoil energy of approximately $248$ keV, in a region where the expected background is low. Interpreting this event as a DM signal requires an interaction between DM and nucleons ($N$). Within the EFT framework considering Dirac fermionic DM ($\chi$), there are 20 possible effective Lagrangians, as listed in Ref.\,\cite{Anand:2013yka}.

Generically, these Lagrangians can be written in the form
\begin{eqnarray}
 &\mathcal{L}_i^{s/v} = \frac{d_i^{s/v}}{\Lambda^2}  \bar{\chi} \,\Gamma_\chi \, \chi \, \bar{N} \, \Gamma_N N, \\
 &\Gamma_{\chi/N} = \{ \mathbb{1}, \gamma_5, \gamma^\mu, \gamma^\mu \gamma_5, \sigma^{\mu \nu}, \sigma^{\mu \nu} \gamma_5, \sigma^{\mu \nu}\frac{q_\nu}{m_M}, \sigma^{\mu \nu}\frac{q_\nu}{m_M} \gamma_5 \} \nonumber
 \label{eq:Lagrangian}
\end{eqnarray}
where $q$ denotes the momentum transfer and $N$ can be either a proton or a neutron. Among all possible combinations of the DM and nucleon bilinears constructed using $\Gamma_{\chi/N}$, we focus on the Lagrangians for which the local significance of the LZ event  $\geq 3\sigma$ at least for one of the DM masses in the Tab. S7 of Ref.\,\cite{LZ:2026axp}. A list of these Lagrangians is provided in Ref.~\cite{Bose:2026szs}. The superscripts $s$ and $v$ in Eq.\,\eqref{eq:Lagrangian} denote isoscalar and isovector couplings, respectively. The cutoff scale is denoted by $\Lambda$. In our direct detection analysis, we fix $\Lambda = v = 246.2$ GeV and $m_M = 0.938$ GeV, following the convention adopted in Ref.\,\cite{LZ:2026axp}. With this choice, the resulting coefficients $d_i^{s/v}$ can be interpreted as proxies for the energy scale probed by direct detection experiments, with the corresponding normalization scales set by $v$ and, where applicable, $m_M$.

Starting from the relativistic Lagrangians in Eq.\,\eqref{eq:Lagrangian}, one can perform a non-relativistic expansion of the DM and nucleon spinors and retain the leading terms in the momentum transfer. This procedure leads to the non-relativistic effective operators $\mathcal{O}_j$ \cite{Fitzpatrick:2012ix}. As required by non-relativistic interactions, these operators are constructed from Galilean-invariant and Hermitian building blocks $i\vec{q}/m_N$, $\vec{v}^{\perp}$, $\vec{S}\chi$, and $\vec{S}_N$, where $\vec{v}^{\perp}$ denotes the relative DM-nucleon velocity perpendicular to the momentum transfer. At the order relevant for our analysis, these include the standard set of 14 non-relativistic operators. Parametrically, the Lagrangian can be written as
\begin{equation}
\mathcal{L}_i \to \mathcal{L}_{\rm eff}= \sum_j \left( c^p_j  \mathcal{O}^p_j + c^n_j \mathcal{O}^n_j \right),  	
\label{eq:Leff}
\end{equation}
where the superscripts $p$ and $n$ denote proton and neutron couplings, respectively. An isoscalar interaction corresponds to $c_j^p=c_j^n$, whereas an isovector interaction corresponds to $c_j^p=-c_j^n$. The sum over $j$ runs over all non-relativistic operators contributing to the interaction.

The DM-nucleus scattering amplitude is obtained by summing the corresponding DM-nucleon operators over all nucleons in the target nucleus. For a scattering process characterized by the momentum transfer $\vec q$, it is convenient to work in Fourier space, where the nucleon operators are expressed as Fourier transforms of the corresponding coordinate-space operators. The resulting momentum space operators can then be decomposed into multipoles in terms of relevant Bessel functions. This multipole expansion allows the nuclear matrix elements to be organized in terms of the relevant nuclear response functions. The nuclear matrix elements can then be organized into a set of nuclear response functions. The resulting DM-nucleus transition probability can be written as
\begin{eqnarray}
	T(v,q) =  \frac{4 \pi}{2j_T + 1} \sum_{k=1}^8 \sum_{x=s,v} \sum_{x^\prime=s,v}  R_k^{x,x'}(v^2, q^2) W_k^{x,x'}(q^2) 	
	\label{eq:trans_prob}
\end{eqnarray}
where $j_T$ denotes the spin of the target nucleus. The DM response functions $R_k^{x,x'}(v^2,q^2)$ contain the dependence on the DM particle physics and the non-relativistic EFT couplings, while the nuclear response functions $W_k^{x,x'}(q^2)$ encode the nuclear-structure information. For elastic scattering and as the nuclear ground state is approximately charge-parity and parity even, the possible nuclear responses reduce to eight independent types, including the interference terms. In our analysis, we adopt the updated GCN nuclear structure calculations implemented in \texttt{dmscatter}\,\cite{Gorton:2022eed}.

The differential scattering cross section is directly related to the scattering transition probabilities $T(v, q(E_r))$:
\begin{equation}
	\frac{d\sigma}{dE_r}(v, E_r) = 2m_T \frac{d\sigma}{dq^2}(v, q) = \frac{2m_T}{4\pi v^2} T(v, q). \quad
\end{equation}
The momentum transfer $q$ is directly related to the recoil energy by $q^2 = 2m_T E_r$, where $m_T$ is the mass of the target nucleus and $E_r$ is the recoil energy.

The differential rate is given by 
\begin{equation}
	\frac{dR}{dE_r} = \frac{1}{m_T} \frac{\rho_{\mathrm{\chi}}}{m_{\mathrm{\chi}}} \int_{v_{\min}} d^3v \, v \, f_{\mathrm{\chi}}(\vec{v} + \vec{v}_{\mathrm{E}}) \frac{d\sigma}{dE_r} (v,E_r)
	\label{eq:diff_rate}
\end{equation}
The local DM density is $\rho_{\chi}=0.3$ GeV/cm$^3$\,\cite{deSalas:2019pee}, $m_{\chi}$ is the DM mass, and $ v_{\min} $ is the minimum DM speed required to produce the recoil energy $E_r$. The DM velocity distribution is assumed to be the Maxwell-Boltzmann distribution with a cut-off at the DM escape velocity $v_{\rm esc}$. These astrophysical parameters are fixed to a value as recommended in Ref.\,\cite{Baxter:2021pqo}. 

In practice, we use the code \texttt{dmscatter} to calculate the event rate in Eq.\,\eqref{eq:diff_rate}, expressing the Lagrangians in terms of linear combinations of non-relativistic operators following Ref.\,\cite{Anand:2013yka}. We also account for the LZ detector's energy resolution, parameterized as $\Delta E_r = 1.46 \sqrt{E_r/\text{keV}}\,\text{keV}$, which is inferred from the statistical uncertainty of $23$ keV at a recoil energy of $248$ keV \cite{LZ:2026axp,McCabe:2026crm}. Incorporating the detector exposure ($\mathcal{N}$) of $2.84$ ton-years, the detection efficiency $\epsilon(E_r)$, and the Gaussian energy-resolution function, we obtain the differential event rate as a function of the reconstructed recoil energy $E_{\rm rec}$
\begin{equation}
	\frac{dR}{dE_{\text{rec}}} = \mathcal{N} \int \frac{e^{-(E_{\text{rec}}-E_r)^2/2\Delta E_r^2}}{\sqrt{2\pi}\Delta E_r} \epsilon(E_r) \frac{dR}{dE_r} \, dE_r
	\label{eq:rec_diff_rate}
\end{equation}
Using Eq.\,\eqref{eq:rec_diff_rate} we followed the methodology of Ref.\,\cite{Bose:2026szs} to obtain best fit parameter space. The resulting regions, bounded by the red dashed lines in Figs.\,\ref{fig:L_10_v}, \ref{fig:bounds_isoscalar} and \ref{fig:bounds_isovector}, are referred to as the LZ band.

\section{Collider analysis}
\label{sec:coll_ana}

\begin{figure*}[t]
\begin{center}
\includegraphics[angle=0.0,width=0.31\textwidth]{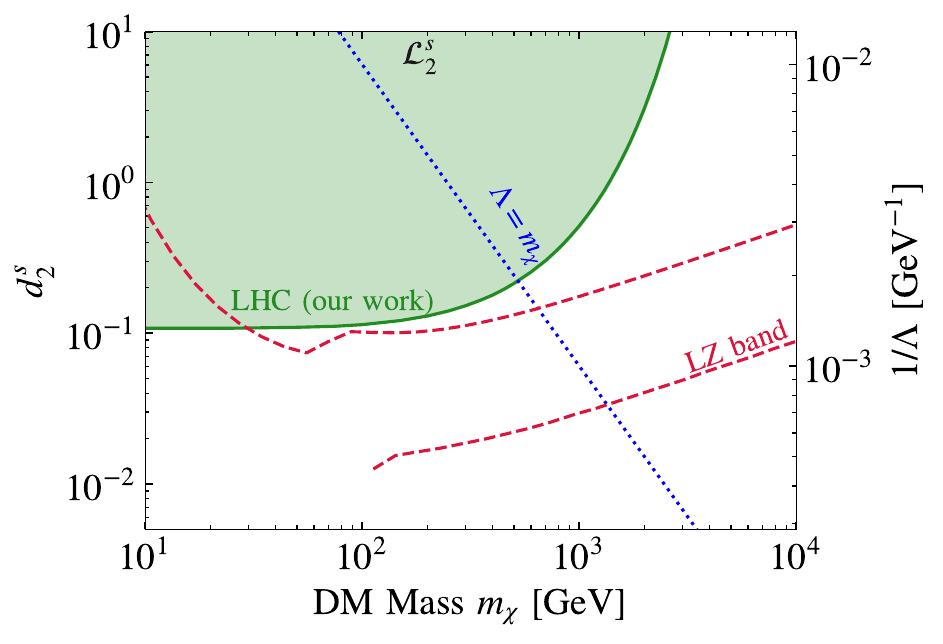}
\includegraphics[angle=0.0,width=0.31\textwidth]{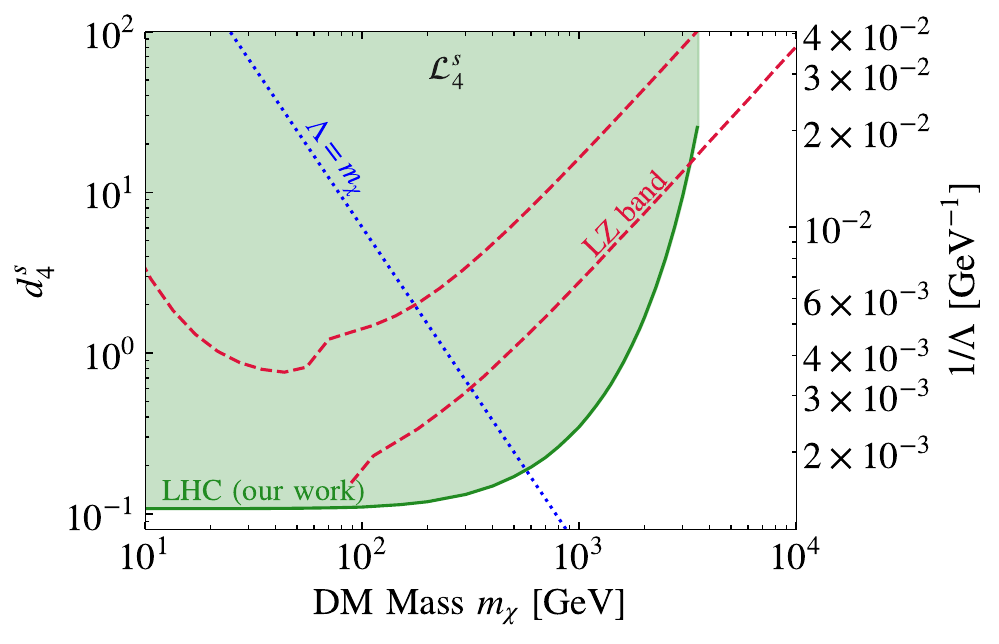}
\includegraphics[angle=0.0,width=0.31\textwidth]{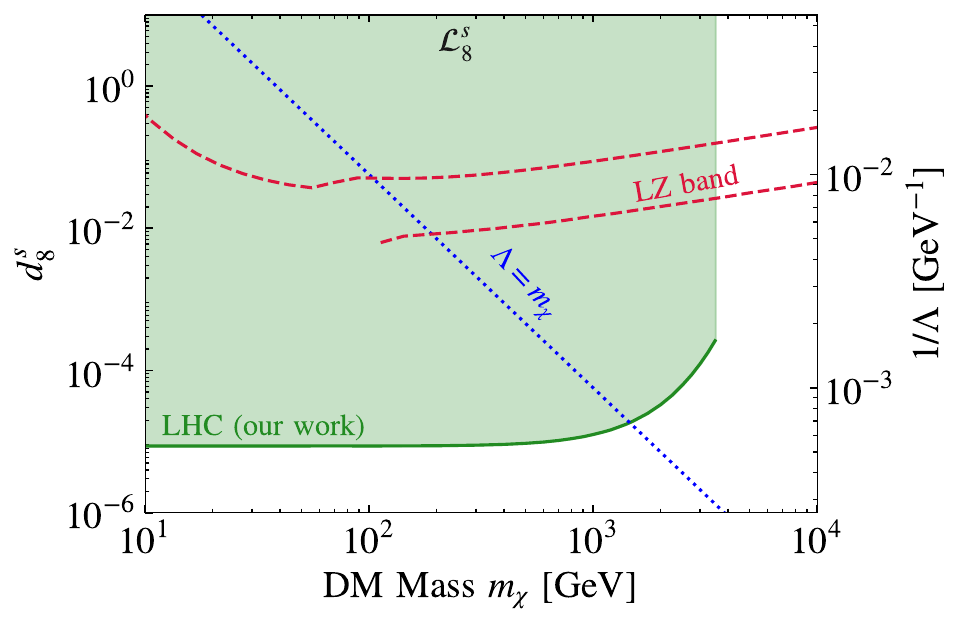}\\
\includegraphics[angle=0.0,width=0.31\textwidth]{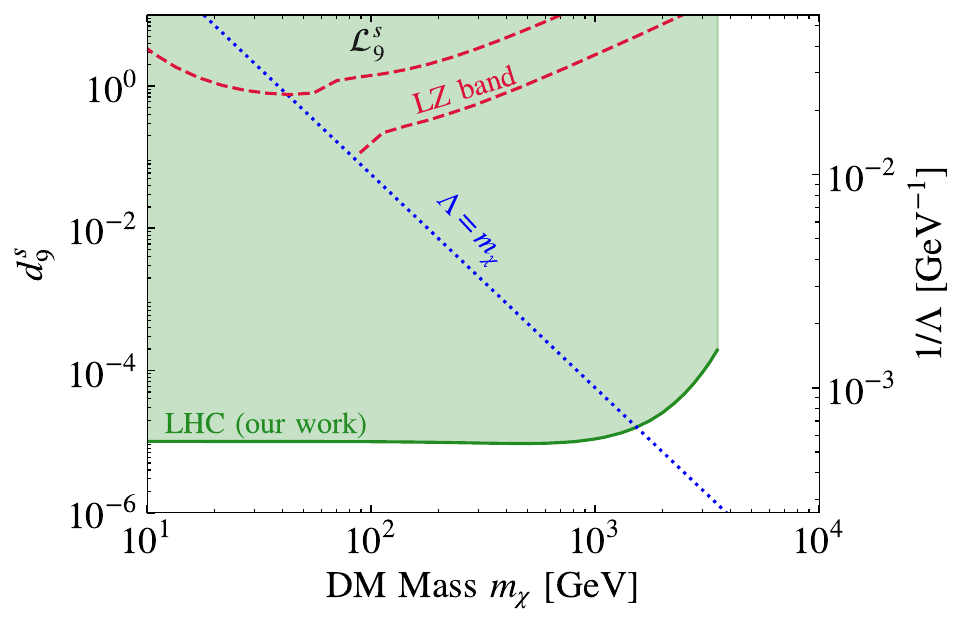}
\includegraphics[angle=0.0,width=0.31\textwidth]{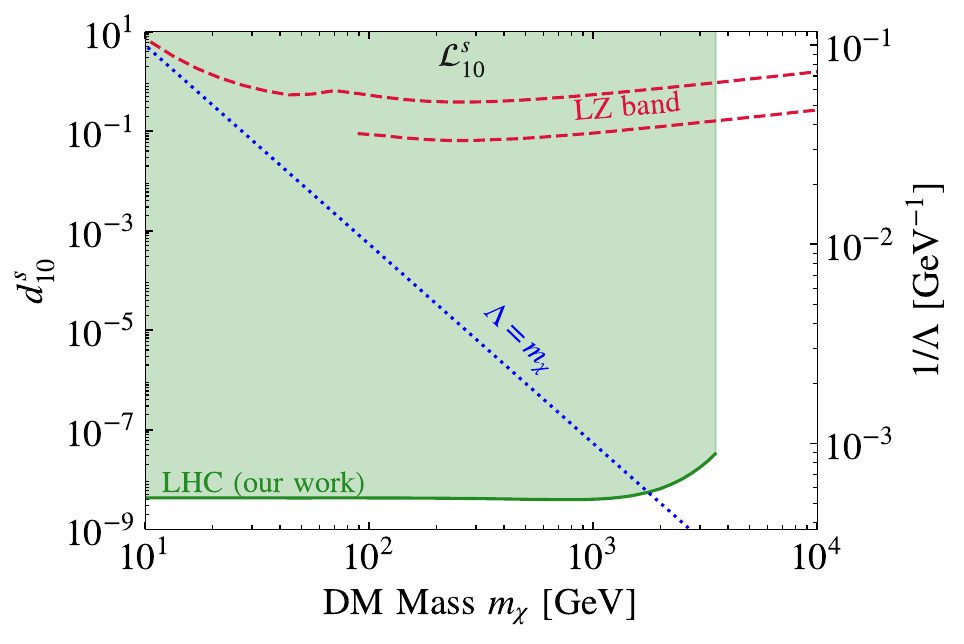}
\includegraphics[angle=0.0,width=0.31\textwidth]{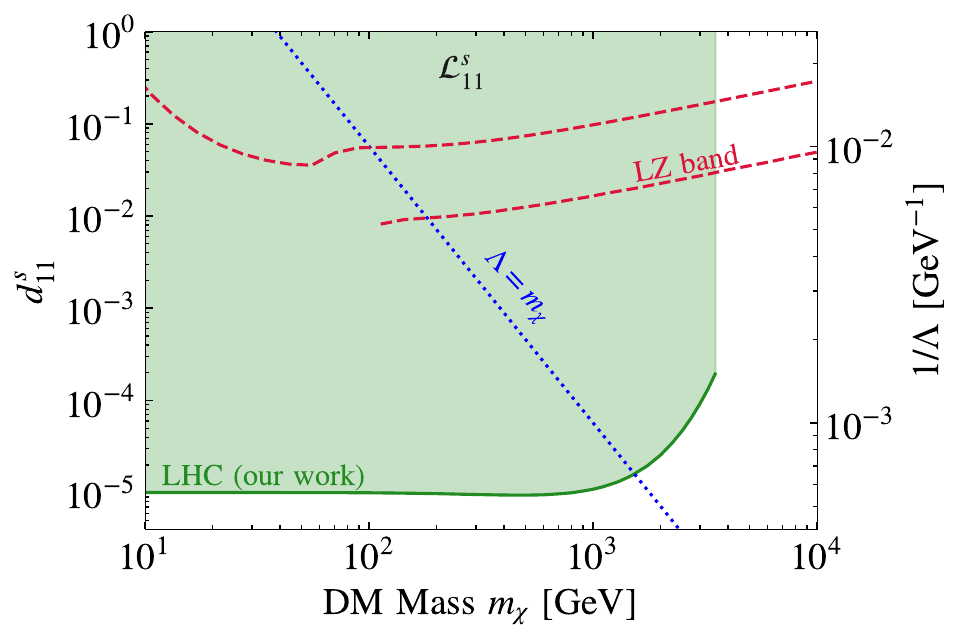}\\
\includegraphics[angle=0.0,width=0.31\textwidth]{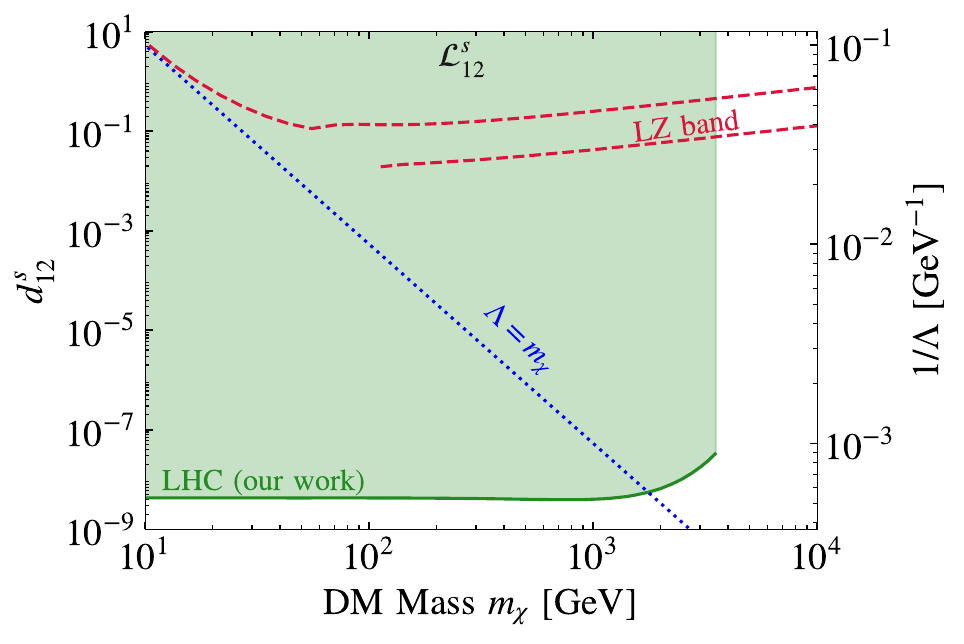}
\includegraphics[angle=0.0,width=0.31\textwidth]{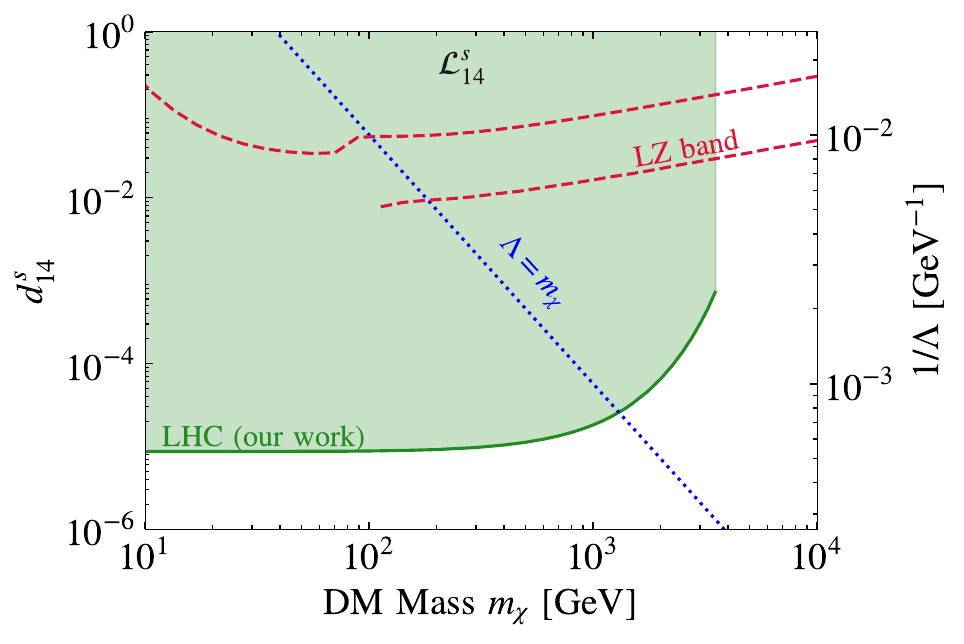}
\includegraphics[angle=0.0,width=0.31\textwidth]{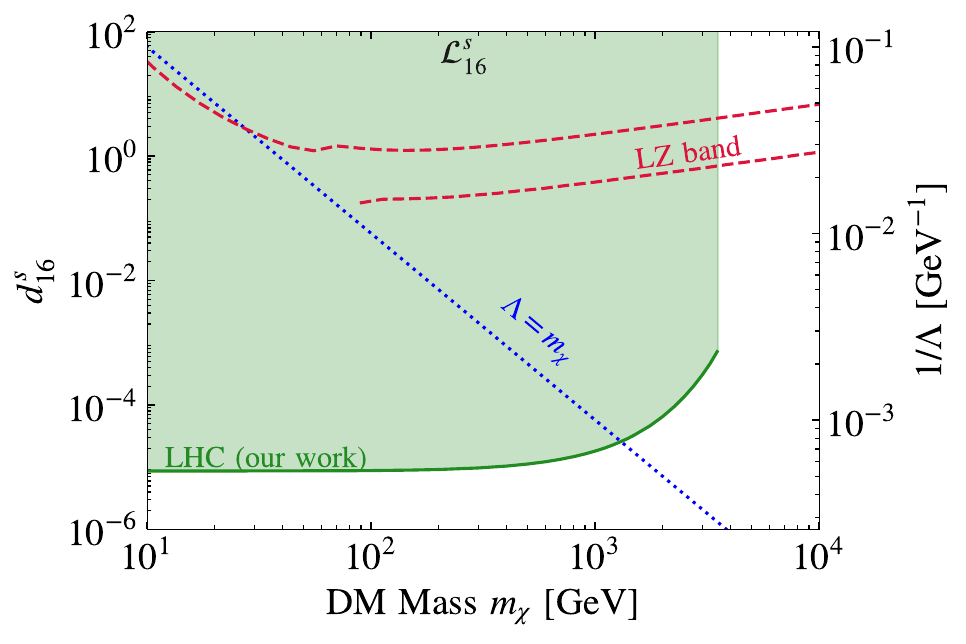}\\
\includegraphics[angle=0.0,width=0.31\textwidth]{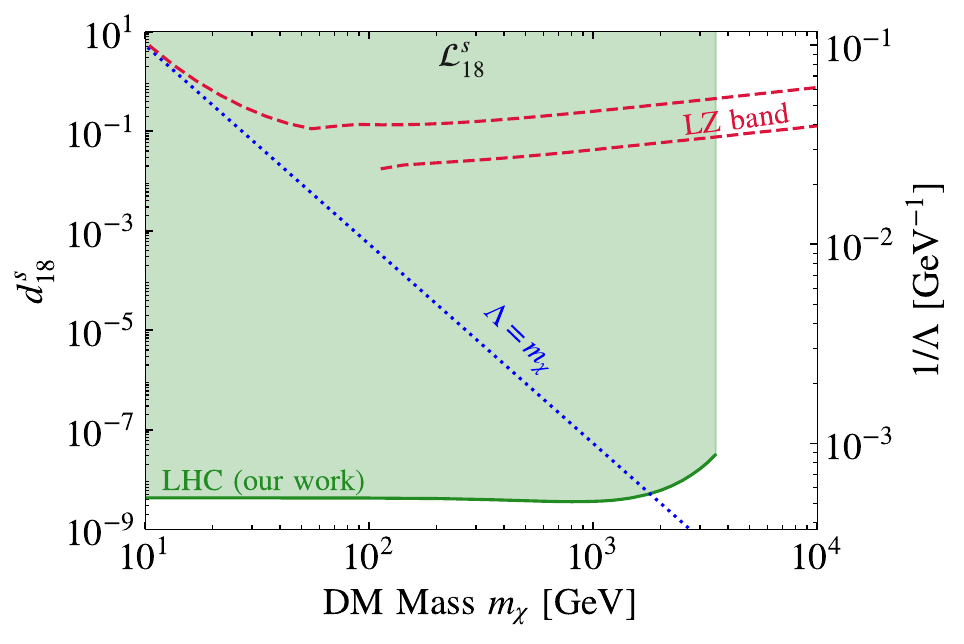}
\includegraphics[angle=0.0,width=0.31\textwidth]{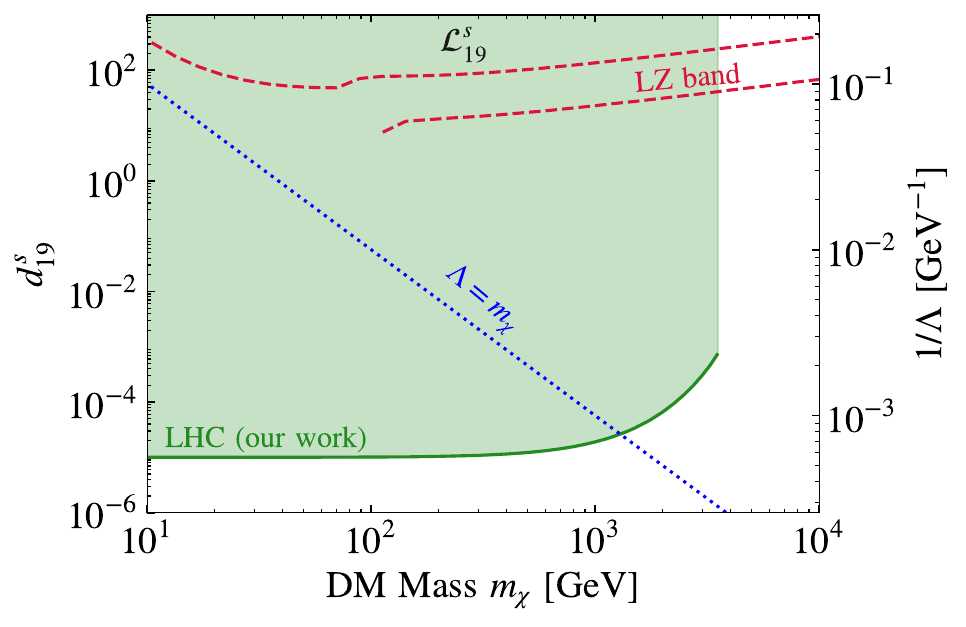}
\includegraphics[angle=0.0,width=0.31\textwidth]{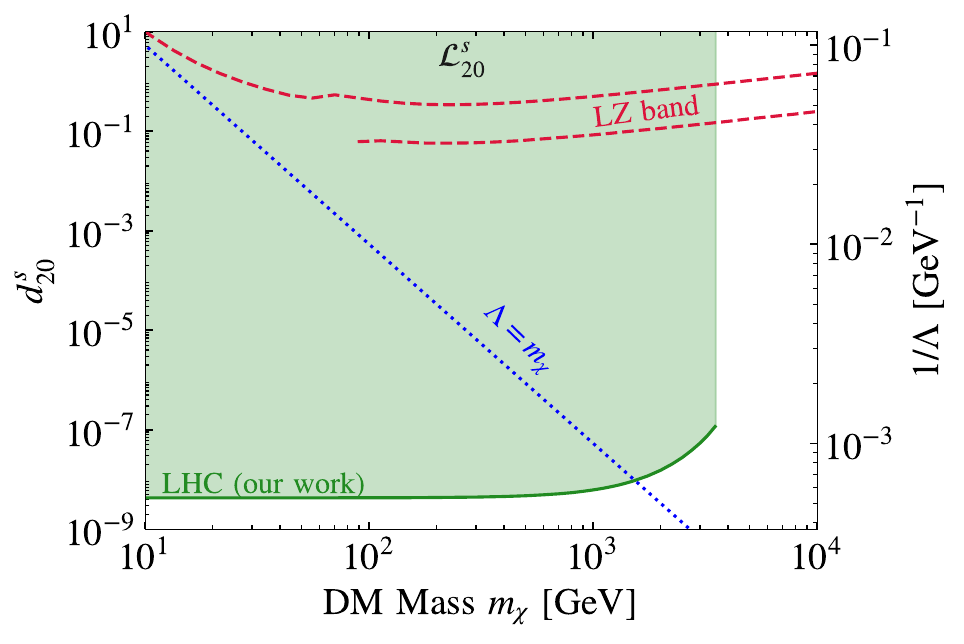}  
    \caption{Analogous to Fig.~\ref{fig:L_10_v}, but for various isoscalar interactions  mentioned in Tab.~\ref{tab:lz_lhc}.}
	\label{fig:bounds_isoscalar}
\end{center}	
\end{figure*}

In this section, we investigate the constraints on DM-nucleon effective interactions at the 13 TeV LHC, focusing on the monojet $+\mET$ signature. However, before presenting the details of our analysis, we clarify the conventions adopted in the collider study and the matching procedure used to implement the DM–quark interactions. Unlike in the direct-detection analysis, where the coefficients $d_i^{s/v}$ are determined by the chosen normalization convention, in the collider analysis we set $d_i^{s/v}=1$ and $m_M=\Lambda$. Consequently, $\Lambda$ serves as a new physics parameter alongside DM mass.
For the collider analysis, we have implemented the relevant DM–quark interactions in \texttt{FeynRules}\,\cite{Alloul:2013bka}. The quark-level couplings are chosen to reproduce the desired isoscalar and isovector coupling structures at the nucleon level. For the scalar, pseudoscalar, vector, and axial-vector nucleon currents, we perform the matching between the quark-level and nucleon-level interactions following the prescription of Ref.\,\cite{Cirelli:2013ufw, DelNobile:2021wmp}. For interactions involving nucleon tensor currents of the form $\sigma^{\mu\nu}q_\nu$ and $\sigma^{\mu\nu}q_\nu\gamma_5$, we determine the corresponding quark-level coefficients using the nucleon matrix elements of the quark tensor current. For example, the Lagrangian $\mathcal L_{10}$ can, in principle, be generated through electromagnetic interactions involving derivative couplings. However, such constructions will also induce vector currents. Since our objective is to investigate collider constraints on individual Lagrangians while avoiding the introduction of additional interactions, we adopt a tensor current based implementation at the quark level.

\begin{figure*}[t]
\begin{center}
\includegraphics[angle=0.0,width=0.31\textwidth]{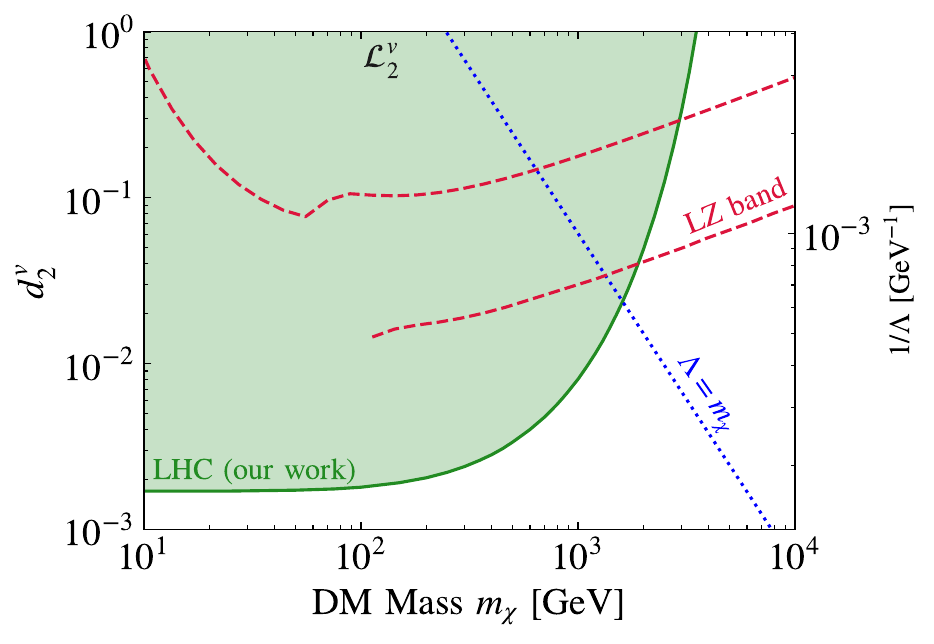}
\includegraphics[angle=0.0,width=0.31\textwidth]{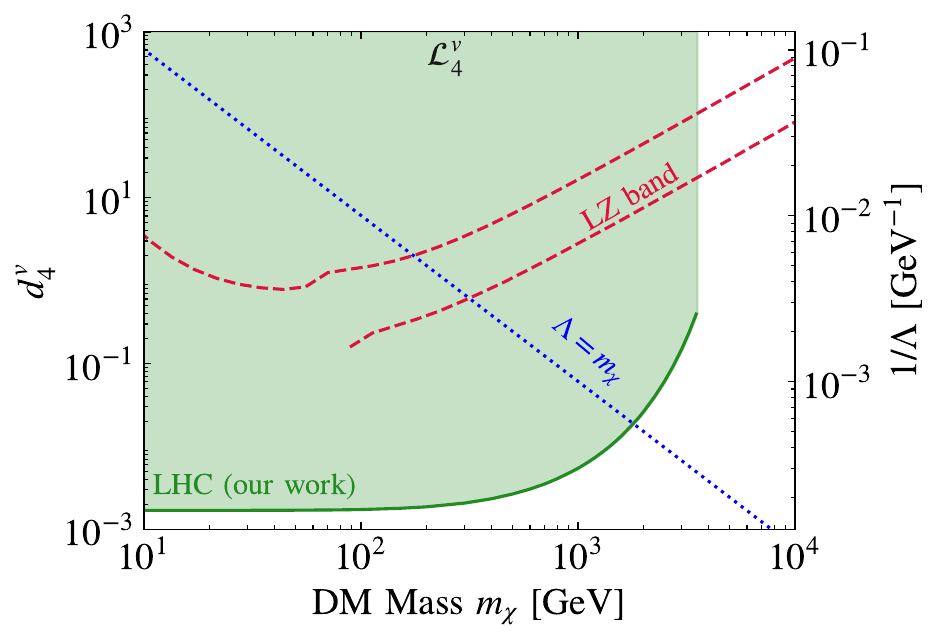}
\includegraphics[angle=0.0,width=0.31\textwidth]{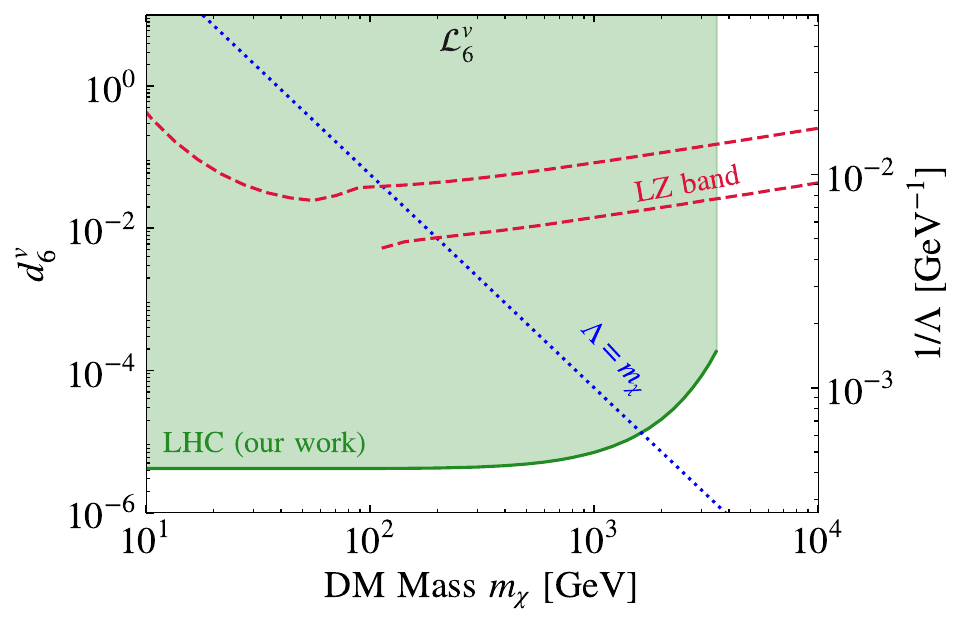}\\
\includegraphics[angle=0.0,width=0.31\textwidth]{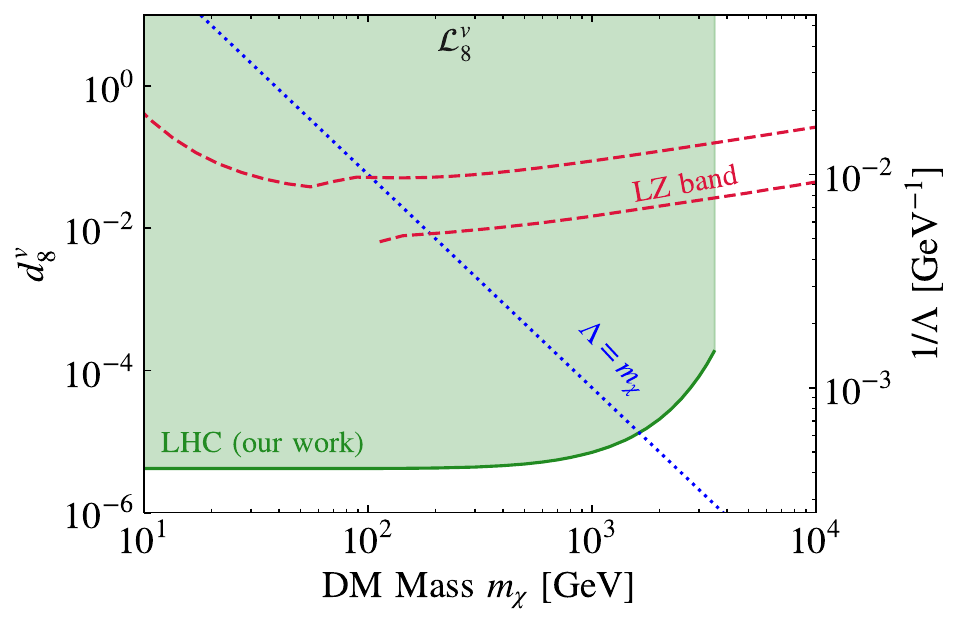}
\includegraphics[angle=0.0,width=0.31\textwidth]{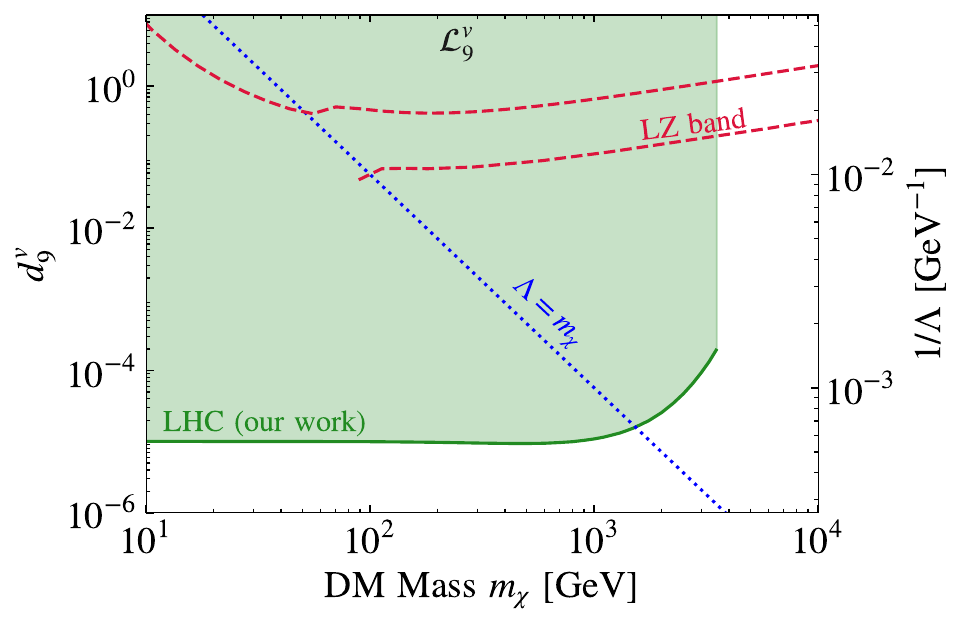}
\includegraphics[angle=0.0,width=0.31\textwidth]{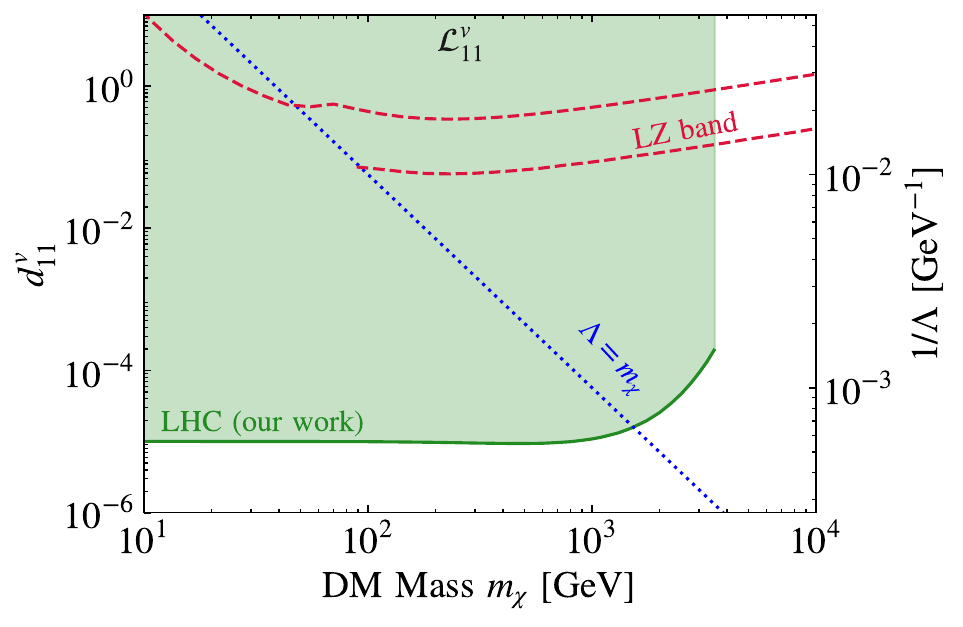}\\
\includegraphics[angle=0.0,width=0.31\textwidth]{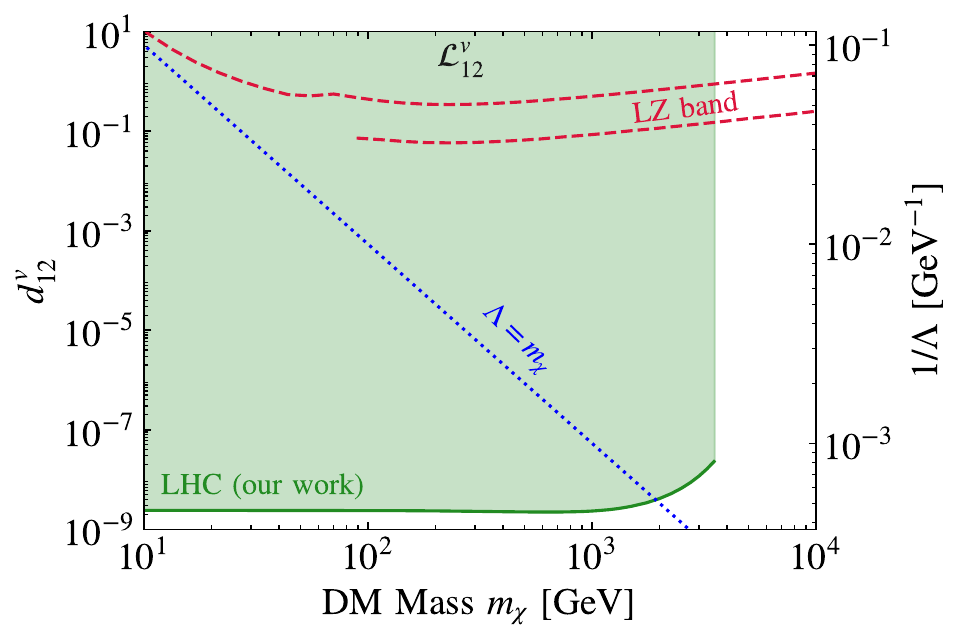}
\includegraphics[angle=0.0,width=0.31\textwidth]{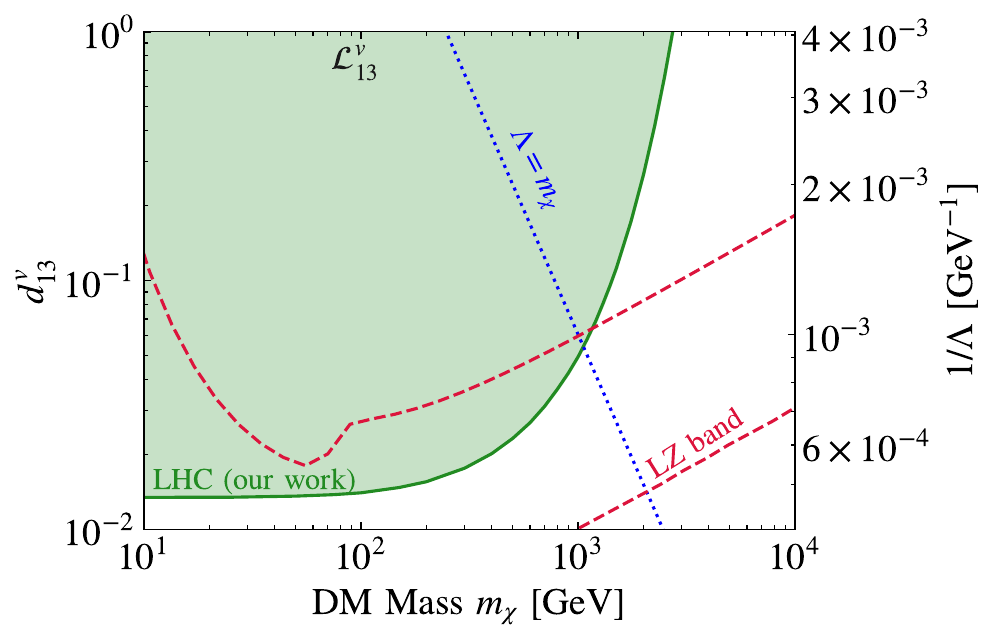}
\includegraphics[angle=0.0,width=0.31\textwidth]{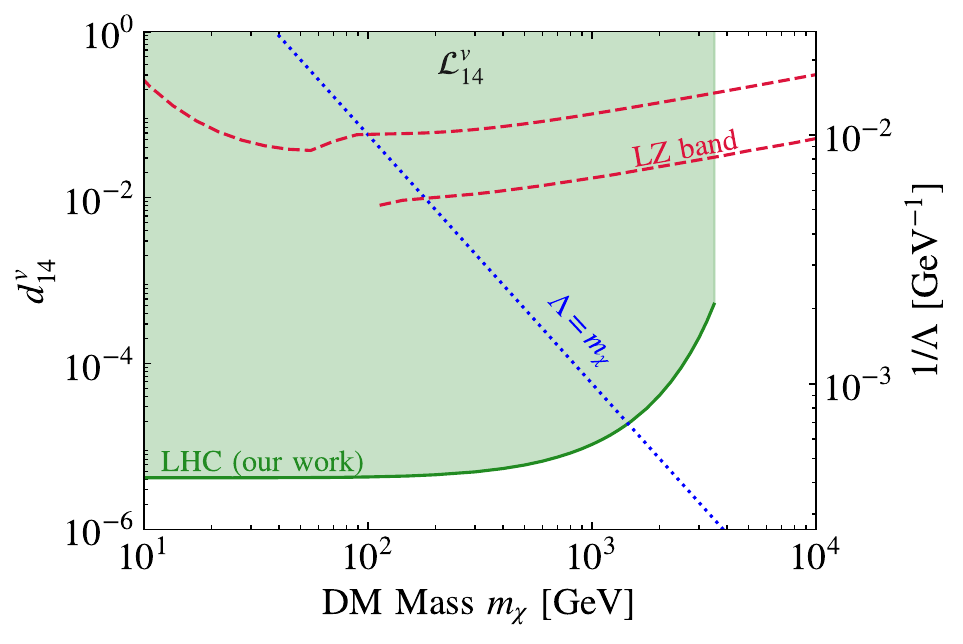}\\
\includegraphics[angle=0.0,width=0.31\textwidth]{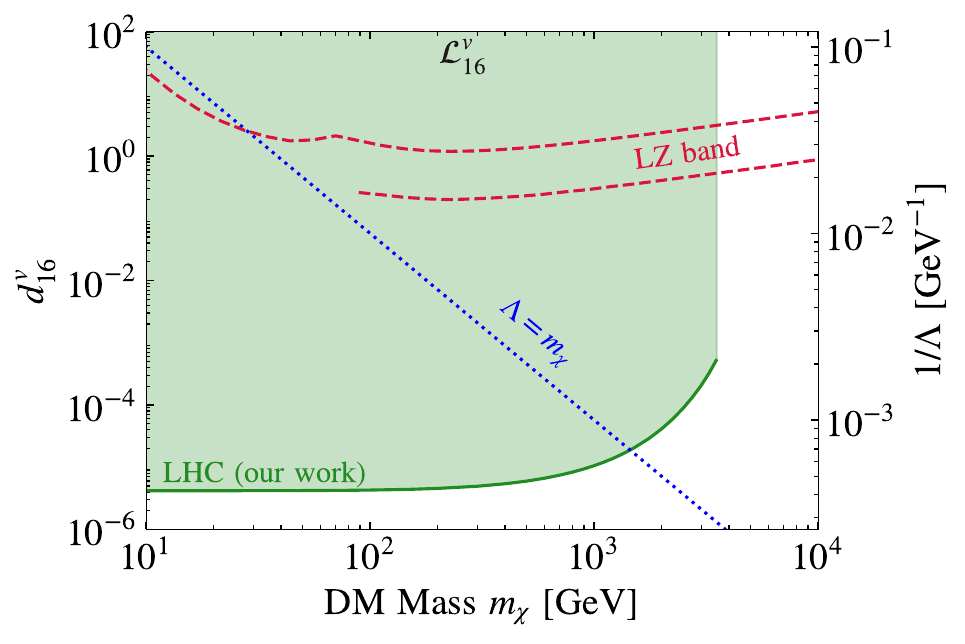}
\includegraphics[angle=0.0,width=0.31\textwidth]{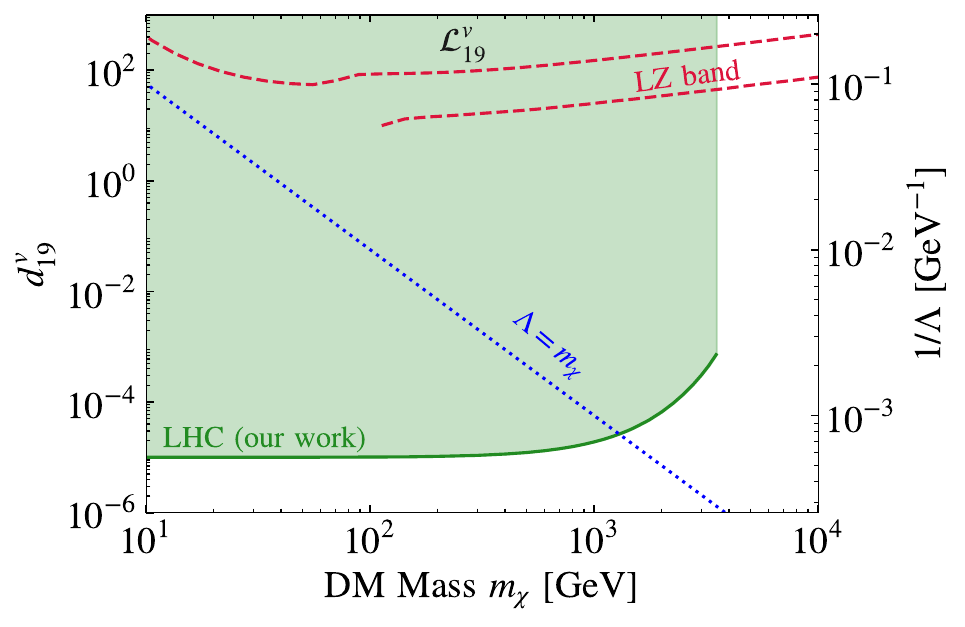}
\includegraphics[angle=0.0,width=0.31\textwidth]{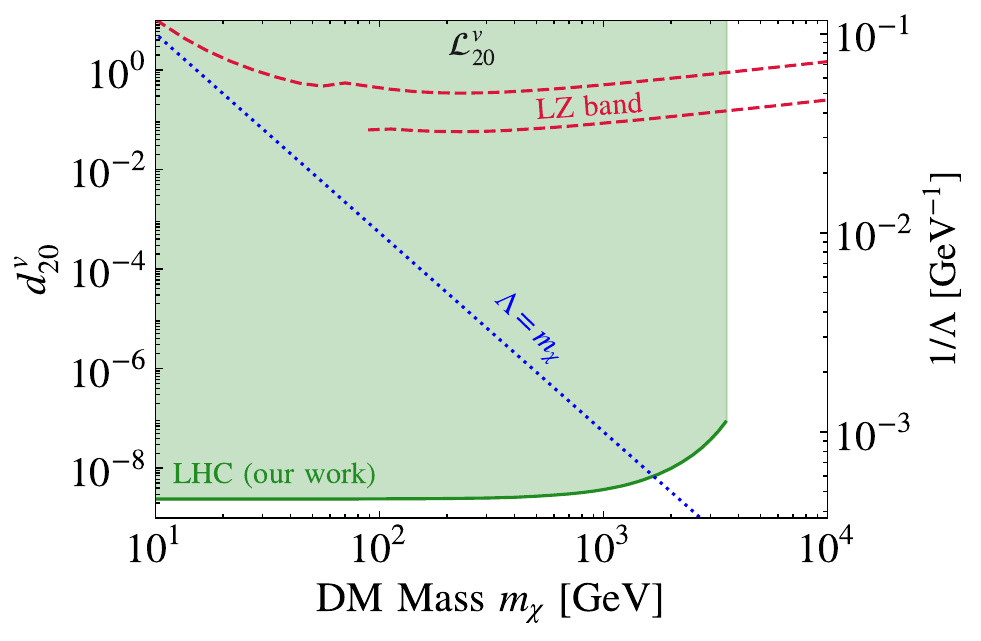}
    \caption{Analogous to Fig.~\ref{fig:L_10_v}, but for various isovector interactions mentioned in Tab.~\ref{tab:lz_lhc}.}
	\label{fig:bounds_isovector}
\end{center}	
\end{figure*}

Having discussed the structure of the required quark couplings, we now turn to the collider analysis. Since the ATLAS Collaboration does not provide direct limits on the effective scale $\Lambda$ for the specific operators considered in this work, we derive the corresponding constraints by recasting the ATLAS search for events with an energetic jet and missing transverse momentum. The signal process of our interest is:
\begin{equation}
pp \to \chi\bar{\chi}+j,
\label{eq:signal}
\end{equation}
where the DM pair escapes the detector without depositing energy, giving rise to missing transverse momentum, while the energetic jet is predominantly generated through initial-state radiation (ISR). The analysis is based on the full Run-II ATLAS dataset collected at a center-of-mass energy of $\sqrt{s}=13$ TeV, corresponding to $\mathfrak{L}_{\rm int}=139~\mathrm{fb}^{-1}$ \cite{ATLAS:2021kxv}. The search focuses on events characterized by a high-transverse-momentum jet accompanied by significant missing transverse momentum. Such a topology provides a sensitive probe of DM pair production, where the visible jet is predominantly generated through initial-state radiation. Final state events are selected with a hard leading jet with $P^j_T>150$ GeV and $|\eta|<2.4$ and a large imbalance in the transverse plane, quantified by the missing transverse momentum, $\mET$. Events with additional charged leptons or photons are vetoed. The resulting event sample is examined in 24 distinct signal regions defined according to $\mET$. Twelve of these are exclusive regions, covering successive ranges from 200 GeV to 1200 GeV, together with an overflow region for events with $\mET>1200$ GeV. The remaining twelve regions are inclusive, each corresponding to a lower $\mET$ threshold that increases from 200 GeV toward the upper end of the experimentally accessible range. The use of both exclusive and inclusive selections allows the analysis to retain sensitivity over a broad range of missing-momentum scales, with the high-$\mET$ regions being particularly useful for probing potential contributions from BSM physics.
\begin{table*}[t]
\centering
\small
\setlength{\tabcolsep}{5pt}
\renewcommand{\arraystretch}{1.25}

\begin{tabular}{|l|c|c|}
\hline
\parbox[c]{0.35\textwidth}{\centering Lagrangian} &
\parbox[c]{0.15\textwidth}{\centering Interaction type} &
\parbox[c]{0.35\textwidth}{\centering LHC exclusion of LZ explanation} \\
\hline

\parbox[c]{0.49\textwidth}{\centering $\displaystyle (\bar\chi\chi)(\bar N i\gamma^5 N)$}
& \rule{0pt}{0.50cm} Isoscalar ($\mathcal L_2^s$) & $\times$ \\
\cline{2-3}
& \rule{0pt}{0.50cm} Isovector ($\mathcal L_2^v$) & $m_{\chi} \lesssim 1600\,$GeV\\
\hline

\parbox[c]{0.49\textwidth}{ \centering $\displaystyle (\bar\chi\gamma^5\chi)(\bar N\gamma^5 N)$}
& \rule{0pt}{0.50cm} Isoscalar ($\mathcal L_4^s$) & $m_{\chi} \lesssim 500\,$GeV \\
\cline{2-3}
& \rule{0pt}{0.50cm} Isovector ($\mathcal L_4^v$) & $m_{\chi} \lesssim 1800\,$GeV \\
\hline

\parbox[c]{0.49\textwidth}{\centering $\displaystyle
(\bar\chi\gamma^\mu\chi)
\left(\bar N i\sigma_{\mu\nu}\frac{q^\nu}{m_M}N\right)$}
& \rule{0pt}{0.50cm} Isovector ($\mathcal L_6^v$) & $m_{\chi} \lesssim 1600\,$GeV \\
\hline

\parbox[c]{0.49\textwidth}{\centering $\displaystyle
(\bar\chi\gamma^\mu\chi)
\left(\bar N i\sigma_{\mu\nu}\frac{q_\nu}{m_M}\gamma^5N\right)$}
&\rule{0pt}{0.50cm}  Isoscalar ($\mathcal L_8^s$) & $m_{\chi} \lesssim 1450\,$GeV \\
\cline{2-3}
& \rule{0pt}{0.50cm} Isovector ($\mathcal L_8^v$) & $m_{\chi} \lesssim 1600\,$GeV \\
\hline

\parbox[c]{0.49\textwidth}{\centering $\displaystyle
\left(\bar\chi i\sigma^{\mu\nu}\frac{q_\nu}{m_M}\chi\right)
(\bar N\gamma_\mu N)$}
& \rule{0pt}{0.50cm} Isoscalar ($\mathcal L_9^s$) & $m_{\chi} \lesssim 1500\,$GeV \\
\cline{2-3}
& \rule{0pt}{0.50cm} Isovector ($\mathcal L_9^v$) & $m_{\chi} \lesssim 1500\,$GeV \\
\hline

\parbox[c]{0.49\textwidth}{\centering $\displaystyle
\left(\bar\chi i\sigma^{\mu\alpha}\frac{q_\alpha}{m_M}\chi\right)
\left(\bar N i\sigma_{\mu\beta}\frac{q^\beta}{m_M}N\right)$}
& \rule{0pt}{0.50cm} Isoscalar ($\mathcal L_{10}^s$) & $m_{\chi} \lesssim 1800\,$GeV \\
\cline{2-3}
& \rule{0pt}{0.50cm} Isovector ($\mathcal L_{10}^v$) & $m_{\chi} \lesssim 1900\,$GeV \\
\hline

\parbox[c]{0.49\textwidth}{\centering $\displaystyle
\left(\bar\chi i\sigma^{\mu\nu}\frac{q_\nu}{m_M}\chi\right)
(\bar N\gamma_\mu\gamma^5N)$}
& \rule{0pt}{0.50cm} Isoscalar ($\mathcal L_{11}^s$) &  $m_{\chi} \lesssim 1500\,$GeV  \\
\cline{2-3}
& \rule{0pt}{0.50cm} Isovector ($\mathcal L_{11}^v$) &  $m_{\chi} \lesssim 1500\,$GeV \\
\hline

\parbox[c]{0.49\textwidth}{\centering $\displaystyle
\left(i\bar\chi i\sigma^{\mu\nu}\frac{q_\nu}{m_M}\chi\right)
\left(\bar N i\sigma_{\mu\alpha}\frac{q^\alpha}{m_N}\gamma^5N\right)$}
& \rule{0pt}{0.50cm} Isoscalar ($\mathcal L_{12}^s$) & $m_{\chi} \lesssim 1750\,$GeV \\
\cline{2-3}
& \rule{0pt}{0.50cm} Isovector ($\mathcal L_{12}^v$) & $m_{\chi} \lesssim 1900\,$GeV\\
\hline

\parbox[c]{0.49\textwidth}{\centering $\displaystyle
(\bar\chi\gamma^\mu\gamma^5\chi)(\bar N\gamma_\mu N)$}
& \rule{0pt}{0.50cm} Isovector ($\mathcal L_{13}^v$) & $m_{\chi} \lesssim 1050\,$GeV (partially)\\
\hline

\parbox[c]{0.49\textwidth}{\centering $\displaystyle
(\bar\chi\gamma^\mu\gamma^5\chi)
\left(\bar N i\sigma_{\mu\nu}\frac{q^\nu}{m_M}N\right)$}
& \rule{0pt}{0.50cm} Isoscalar ($\mathcal L_{14}^s$) & $m_{\chi} \lesssim 1300\,$GeV  \\
\cline{2-3}
&\rule{0pt}{0.50cm} Isovector ($\mathcal L_{14}^v$) & $m_{\chi} \lesssim 1450\,$GeV \\
\hline

\parbox[c]{0.49\textwidth}{\centering $\displaystyle
(i\bar\chi\gamma^\mu\gamma^5\chi)
\left(\bar N i\sigma_{\mu\nu}\frac{q^\nu}{m_M}\gamma^5N\right)$}
& \rule{0pt}{0.50cm} Isoscalar ($\mathcal L_{16}^s$) & $m_{\chi} \lesssim 1300\,$GeV \\
\cline{2-3}
& \rule{0pt}{0.50cm} Isovector ($\mathcal L_{16}^v$) & $m_{\chi} \lesssim 1450\,$GeV \\
\hline

\parbox[c]{0.49\textwidth}{\centering $\displaystyle
\left(i\bar\chi i\sigma^{\mu\nu}\frac{q_\nu}{m_N}\gamma^5\chi\right)
\left(\bar N i\sigma_{\mu\alpha}\frac{q^\alpha}{m_M}N\right)$}
& \rule{0pt}{0.50cm} Isoscalar ($\mathcal L_{18}^s$) & $m_{\chi} \lesssim 1850\,$GeV \\
\hline

\parbox[c]{0.49\textwidth}{\centering $\displaystyle
\left(i\bar\chi i\sigma^{\mu\nu}\frac{q_\nu}{m_M}\gamma^5\chi\right)
(\bar N\gamma_\mu\gamma^5N)$}
& \rule{0pt}{0.50cm} Isoscalar ($\mathcal L_{19}^s$) & $m_{\chi} \lesssim 1250\,$GeV \\
\cline{2-3}
& \rule{0pt}{0.50cm} Isovector ($\mathcal L_{19}^v$) & $m_{\chi} \lesssim 1300\,$GeV\\
\hline

\parbox[c]{0.49\textwidth}{\centering $\displaystyle
\left(i\bar\chi i\sigma^{\mu\nu}\frac{q_\nu}{m_M}\gamma^5\chi\right)
\left(\bar N i\sigma_{\mu\alpha}\frac{q^\alpha}{m_M}\gamma^5N\right)$}
& \rule{0pt}{0.50cm} Isoscalar ($\mathcal L_{20}^s$) & $m_{\chi} \lesssim 1500\,$GeV \\
\cline{2-3}
& \rule{0pt}{0.50cm} Isovector ($\mathcal L_{20}^v$) & $m_{\chi} \lesssim 1700\,$GeV \\
\hline

\end{tabular}
\caption{Effective Lagrangians considered in this analysis and the corresponding DM mass regions of the LZ interpretation excluded by LHC constraints. A $\times$ indicates that the LZ-preferred DM parameter space remains consistent with the LHC constraints over the entire DM mass range considered. We emphasize that different normalizations are adopted in the direct-detection and collider analyses. For direct detection, we fix $\Lambda=v=246$ GeV and $m_M=0.938$ GeV, while varying $d_i^{s/v}$. For the collider analysis, we set $d_i^{s/v}=1$ and $m_M=\Lambda$, while varying $\Lambda$. We note that the validity of the EFT description in the context of LHC in some of the LZ excluded regions require attention\,\cite{Roy:2025inq, Guchait:2020wqn, Busoni:2013lha}}.
\label{tab:lz_lhc}
\end{table*}

To derive constraints on the NP parameters ($m_{\chi}, \Lambda$), we construct a $\chi^2$ function defined as
\begin{equation}
\chi^2 = \sum_{\mET~\text{bins}} \frac{\left(\mathcal{N}^{\rm obs} - \mathcal{N}^{\rm th}(m_{\chi}, \Lambda)\right)^2}{(\Delta \mathcal{N})^2},
\label{eq:chisqr}
\end{equation}
where $\mathcal{N}^{\rm obs}$ denotes the observed number of events in each bin, $\mathcal{N}^{\rm th}(m_{\chi}, \Lambda)$ is the corresponding theoretical prediction as a function of the model parameters, and $\Delta \mathcal{N}$ represents the statistical uncertainty in each bin. The sum runs over all bins of the missing transverse energy ($\mET$) distribution. We generate the signal and the SM background events in $\tt{MadGraph5@NLO} $~\cite{Alwall:2014hca}. The events are then passed to $\tt{PYTHIA8} $\cite{Sjostrand:2014zea} for hadronization, and finally, detector simulation is performed using $\tt{DELPHESv3}$ \cite{deFavereau:2013fsa}. Using the $\chi^2$ statistic defined in Eq.\,\eqref{eq:chisqr}, we derive the constraints on the effective scale $\Lambda$ at the 95\% confidence level.

For the isoscalar scenario, the constraints on $1/\Lambda$ (equivalently on $d_i^{s/v}$) as a function of the DM mass are presented in Fig.~\ref{fig:bounds_isoscalar}.
 The $\mathcal{L}_{2,4}$ Lagrangians are dimension-six effective interactions and do not contain any explicit momentum dependence. Consequently, they lead to the weakest constraints among the Lagrangians considered. In contrast, the $\mathcal{L}_{8,9,11,14,16,19}$ Lagrangians are dimension-seven effective interactions and have momentum dependence through either the DM current or the quark/nucleon current, resulting in stronger constraints than those obtained from $\mathcal{L}_{2,4}$ Lagrangians. Among these Lagrangians, the constraints on $\mathcal{L}_{8,14,16}$ Lagrangians are stronger by a factor of a few compared with those on $\mathcal{L}_{11,19}$ operators. This can be attributed to the momentum dependence residing in the quark/nucleon currents for $\mathcal{L}_{8,14,16}$ Lagrangians, which provides a stronger enhancement of the scattering amplitude. Finally, the $\mathcal{L}_{10,12,18,20}$ Lagrangians arise at dimension-eight and exhibit momentum dependence in both the DM and quark/nucleon currents. Consequently, they receive the largest momentum enhancement and yield the strongest constraints among all the operators considered. 
 
 For the isovector scenario, the corresponding constraints are presented in Fig.~\ref{fig:bounds_isovector}. The overall behavior is similar to that observed in the isoscalar case, with the constraints following the same hierarchy according to the dimension and momentum dependence of the operators. Further, for some of the Lagrangians, we obtain stronger constraints on the isovector interactions than on the corresponding isoscalar interactions, owing to differences in the matching between the quark-level and nucleon-level currents. 
 
 In Tab.~\ref{tab:lz_lhc}, we summarize our results. In particular, the last column indicates the DM mass range disfavored by the LHC constraints derived in this work. In quoting these mass ranges, we also account for the indicative EFT validity limit, $\Lambda=m_\chi$. The symbol `$\times$' denotes the case in which the LHC constraints are weak to exclude any part of the DM parameter space required to explain the LZ event. We re-emphasize the different normalization conventions adopted for the direct-detection and collider analyses: for direct detection, we fix $\Lambda=v=246$ GeV and $m_M=0.938$ GeV and vary $d_i^{s/v}$, whereas for the collider analysis, we set $d_i^{s/v}=1$ and $m_M=\Lambda$ and vary $\Lambda$. The results presented in Figs.\,\ref{fig:L_10_v}, \ref{fig:bounds_isoscalar} and \ref{fig:bounds_isovector} with $d_i^{s/v}$ on the left $y$-axis and $1/\Lambda$ on the right $y$-axis, with the appropriate normalization taken into account. In closing, we note that, for most of the considered Lagrangians, DM interpretation with masses below a few TeV is disfavored by current LHC constraints. This holds under the assumption that the EFT provides a valid description of the underlying theory, for some interactions, this validity has been shown up to $\Lambda \sim $ a few GeV\,\cite{Roy:2025inq, Guchait:2020wqn}.

\section{Conclusion}
\label{sec:conclusion}

The LZ DM direct detection experiment has reported an event at a recoil energy of $\sim 250\,\mathrm{keV}$ in its extended recoil-energy search, with a global significance of $2.6\sigma$. Naturally, DM provides a potential interpretation of this event through its interactions with nucleons. However, the absence of additional events at lower recoil energies ($\sim\mathrm{keV}$) disfavors conventional elastic scattering with momentum-independent interactions. An elastic interpretation can instead be accommodated by momentum-dependent interactions, which suppress the scattering rate at low momentum transfer while allowing the observed high-energy recoil to be reproduced. In contrast, an inelastic DM interpretation can naturally suppress low-energy recoils through the associated kinematics. In this paper, we focus on the elastic-scattering interpretation.

A variety of elastic DM–nucleon interactions can provide a satisfactory description of the LZ event, with the preferred DM masses typically lying in the TeV range. Given the momentum dependence of these interactions and the corresponding DM mass scale, it is natural to ask whether the regions of parameter space compatible with the LZ event are also consistent with existing collider constraints. 

To address this question, we perform a systematic analysis for Lagrangians having $\geq 3 \sigma$ local significance for the LZ event. For most of the Lagrangian, we find that the current ATLAS monojet search with missing transverse momentum significantly constrains the DM parameter space compatible with the LZ event, pushing the viable DM mass to values above $\sim\mathrm{TeV}$ with the condition that the description of EFT is valid for the considered region. While such constraints are broadly expected for some of the interactions, we emphasize that, to the best of our knowledge, this is the first study of Lagrangians having explicit momentum dependency (such as $\mathcal{L}_{10}$) using current ATLAS data in the context of the LZ event.

Finally, we note that, irrespective of whether future experiments confirm the LZ observation, the region of DM parameter space excluded by the present LHC analysis, particularly for DM masses below $\sim\mathrm{TeV}$, would be challenging to probe with future direct detection experiments alone. This highlights the complementarity of collider searches and direct-detection experiments in testing DM interpretations.

\emph{Acknowledgments --}  We thank Ranjan Laha and Shigeki Matsumoto for useful discussions. TNM thanks Debajit Bose, Nirmal Raj, and Akash Kumar Saha for their help with the LZ fitting procedure and for maintaining a list of useful references. SA acknowledges DAE, Saha Institute of Nuclear Physics (SINP), Kolkata, as his research funding source. NG's work is supported by the Japan Society for the Promotion of Science (JSPS) as part of the JSPS Postdoctoral Program (Standard), grant number: JP24KF0189, and by the World Premier International Research Center Initiative (WPI), MEXT, Japan (Kavli IPMU). NG would also like to thank CD3, Kavli IPMU, for giving access to ChatGPT.  The authors take full responsibility for the scientific contents presented here.

\bibliographystyle{JHEP}
\bibliography{ref.bib}

\providecommand{\href}[2]{#2}\begingroup\raggedright\begin{thebibliography}{100}

\bibitem{LZ:2024zvo}
{\scshape LZ} collaboration, J.~Aalbers et~al., \emph{{Dark Matter Search
  Results from 4.2{\,}{\,}Tonne-Years of Exposure of the LUX-ZEPLIN (LZ)
  Experiment}}, \href{http://dx.doi.org/10.1103/4dyc-z8zf}{\emph{Phys. Rev.
  Lett.} {\bf 135} (2025) 011802},
  [\href{https://arxiv.org/abs/2410.17036}{{\tt 2410.17036}}].

\bibitem{XENON:2025vwd}
{\scshape XENON} collaboration, E.~Aprile et~al., \emph{{WIMP Dark Matter
  Search Using a 3.1 Tonne-Year Exposure of the XENONnT Experiment}},
  \href{http://dx.doi.org/10.1103/msw4-t342}{\emph{Phys. Rev. Lett.} {\bf 135}
  (2025) 221003}, [\href{https://arxiv.org/abs/2502.18005}{{\tt 2502.18005}}].

\bibitem{PandaX:2024qfu}
{\scshape PandaX} collaboration, Z.~Bo et~al., \emph{{Dark Matter Search
  Results from 1.54{\,}{\,}Tonne{\textperiodcentered}Year Exposure of
  PandaX-4T}},
  \href{http://dx.doi.org/10.1103/PhysRevLett.134.011805}{\emph{Phys. Rev.
  Lett.} {\bf 134} (2025) 011805},
  [\href{https://arxiv.org/abs/2408.00664}{{\tt 2408.00664}}].

\bibitem{LZ:2026axp}
{\scshape LZ} collaboration, D.~S. Akerib et~al., \emph{{Search for dark matter
  particle interactions in an extended nuclear recoil energy window with the
  LUX-ZEPLIN (LZ) experiment}},  \href{https://arxiv.org/abs/2609.02823}{{\tt
  2609.02823}}.

\bibitem{Fan:2026kxx}
J.~Fan and M.~Reece, \emph{{Higgsino Above the Sea of Fog}},
  \href{https://arxiv.org/abs/2609.01504}{{\tt 2609.01504}}.

\bibitem{Freese:2026sga}
K.~Freese and D.~P. Theodosopoulos, \emph{{Higgsino Dark Matter Interpretation
  of the LUX-ZEPLIN 248 keV Nuclear-Recoil Event}},
  \href{https://arxiv.org/abs/2609.01583}{{\tt 2609.01583}}.

\bibitem{Wu:2026nhi}
L.~Wu, Y.~Zhang and B.~Zhu, \emph{{TeV Higgsino Dark Matter from LZ Nuclear
  Recoil to Fermi-LAT Gamma Rays}},
  \href{https://arxiv.org/abs/2609.01590}{{\tt 2609.01590}}.

\bibitem{DiMauro:2026ldr}
M.~Di~Mauro, \emph{{Dark Matter at the Kinematic Edge: Interpreting the 248 keV
  LZ Nuclear-Recoil Candidate}},  \href{https://arxiv.org/abs/2609.02608}{{\tt
  2609.02608}}.

\bibitem{Du:2026guj}
X.~Du and F.~Wang, \emph{{TeV Higgsino Interpretation of the LZ High-Recoil
  Event with Intermediate-Scale Electroweak Gauginos}},
  \href{https://arxiv.org/abs/2609.04163}{{\tt 2609.04163}}.

\bibitem{Rodd:2026tyn}
N.~L. Rodd, B.~R. Safdi, T.~R. Slatyer and W.~L. Xu, \emph{{Confronting the
  Higgsino Interpretation of the LZ Event with the High-Energy Sideband}},
  \href{https://arxiv.org/abs/2609.04175}{{\tt 2609.04175}}.

\bibitem{Dent:2026bji}
J.~B. Dent and J.~L. Newstead, \emph{{Exothermic and Endothermic Inelastic Dark
  Matter Interpretations at LZ: Sideband Constraints and Future Prospects}},
  \href{https://arxiv.org/abs/2609.04673}{{\tt 2609.04673}}.

\bibitem{Visinelli:2026kgt}
L.~Visinelli, \emph{{A Peccei-Quinn Origin for Inelastic Electroweak Dark
  Matter after LUX-ZEPLIN}},  \href{https://arxiv.org/abs/2609.02807}{{\tt
  2609.02807}}.

\bibitem{Gu:2026vto}
G.~Gu, L.~Li, S.-S. Tang and Y.~Xu, \emph{{Inelastic from the Other Side: Xenon
  Excitation Signals in Light of the LZ High-Recoil Event}},
  \href{https://arxiv.org/abs/2609.05291}{{\tt 2609.05291}}.

\bibitem{Smirnov:2026aqk}
J.~Smirnov, S.~Griffith and J.~F. Beacom, \emph{{Inelastic Signatures of
  Electroweak Dark Matter}},  \href{https://arxiv.org/abs/2609.04144}{{\tt
  2609.04144}}.

\bibitem{McCabe:2026crm}
C.~McCabe, \emph{{Seasonal dark matter from the LUX-ZEPLIN high-energy event}},
   \href{https://arxiv.org/abs/2609.04181}{{\tt 2609.04181}}.

\bibitem{Chattopadhyay:2026ryw}
U.~Chattopadhyay, D.~Das, R.~Puri and J.~Roy, \emph{{Sub-TeV Singlino Dark
  Matter in light from Sagittarius A$^\ast$ and LUX-ZEPLIN Nuclear-Recoil
  Event}},  \href{https://arxiv.org/abs/2609.02994}{{\tt 2609.02994}}.

\bibitem{deLima:2026shq}
C.~H. de~Lima, \emph{{Exothermic Dark Matter at LZ}},
  \href{https://arxiv.org/abs/2609.05204}{{\tt 2609.05204}}.

\bibitem{Baer:2026fpy}
H.~Baer and V.~Barger, \emph{{Exothermic dark matter and the 248 keV nuclear
  recoil in LUX-ZEPLIN}},  \href{https://arxiv.org/abs/2609.06153}{{\tt
  2609.06153}}.

\bibitem{Bose:2026ndd}
D.~Bose et~al., \emph{{Not so good $\nu$s for Higgsino dark matter as LZ
  excess: stringent limits from Super-Kamiokande and IceCube}},
  \href{https://arxiv.org/abs/2609.07807}{{\tt 2609.07807}}.

\bibitem{Lee:2026wof}
H.~M. Lee, \emph{{Inelastic dark matter and baryon flavor symmetry in light of
  LUX-ZEPLIN (LZ) experiment}},  \href{https://arxiv.org/abs/2609.06171}{{\tt
  2609.06171}}.

\bibitem{Wang:2026ytg}
L.~Wang and Y.~Xiao, \emph{{The Inert Doublet Model of Dark Matter and the
  LUX-ZEPLIN High-Recoil Event}},  \href{https://arxiv.org/abs/2609.06571}{{\tt
  2609.06571}}.

\bibitem{Yang:2026wpb}
M.~Yang, Q.-f. Wu, Y.-L.~S. Tsai and Y.-Z. Fan, \emph{{Multi-Messenger and
  Paleo-Detector Probes of the LZ Dark Matter Signal}},
  \href{https://arxiv.org/abs/2609.06640}{{\tt 2609.06640}}.

\bibitem{Das:2026uyy}
P.~Das, B.~Karmakar, S.~Mahapatra and P.~K. Paul, \emph{{Inelastic
  Self-interacting Dark Matter and LUX-ZEPLIN 248 keV Event in a Dirac Modular
  Inverse Seesaw}},  \href{https://arxiv.org/abs/2609.06825}{{\tt 2609.06825}}.

\bibitem{Barman:2026omh}
B.~Barman, \emph{{Did LZ see modified gravity?}},
  \href{https://arxiv.org/abs/2609.15118}{{\tt 2609.15118}}.

\bibitem{Ahmed:2026qjg}
W.~Ahmed and G.~K. Leontaris, \emph{{A Dark-Dimension Origin of Geometric
  Inelastic Dark Matter: The LUX-ZEPLIN High-Recoil Event and Multi-Target
  Tests}},  \href{https://arxiv.org/abs/2609.07138}{{\tt 2609.07138}}.

\bibitem{DiMauro:2026dqp}
M.~Di~Mauro and H.~Shaikh, \emph{{Solar Capture Tests of Inelastic Dark Matter
  after the LZ High-Recoil Event}},
  \href{https://arxiv.org/abs/2609.06760}{{\tt 2609.06760}}.

\bibitem{Du:2026lpa}
X.-Y. Du, W.~Huang and K.~Xie, \emph{{Pseudo-Dirac Inelastic Dark Matter in the
  Leptophobic $U(1)_B$ Model: Confronting the LUX-ZEPLIN High-Recoil Event with
  Collider Searches}},  \href{https://arxiv.org/abs/2609.07225}{{\tt
  2609.07225}}.

\bibitem{Bandyopadhyay:2026gjw}
D.~Bandyopadhyay, D.~Borah and P.~Borah, \emph{{LZ nuclear recoil event from
  inelastic singlet-doublet scalar dark matter}},
  \href{https://arxiv.org/abs/2609.07451}{{\tt 2609.07451}}.

\bibitem{Borah:2026zwf}
D.~Borah, S.~K. Sahoo, N.~Sahu and S.~Sharma, \emph{{Inelastic Singlet-Doublet
  Fermion Dark Matter in light of the 248 keV LZ event}},
  \href{https://arxiv.org/abs/2609.07800}{{\tt 2609.07800}}.

\bibitem{Bisal:2026khf}
S.~Bisal, J.~Cao and F.~Li, \emph{{Higgsino Dark Matter Interpretation of the
  LZ High-Recoil Event in the GNMSSM with TeV-Scale Gauginos}},
  \href{https://arxiv.org/abs/2609.07811}{{\tt 2609.07811}}.

\bibitem{Cheung:2026byg}
K.~Cheung, S.~K. Kang and R.~Kumar, \emph{{From LUX-ZEPLIN to Colliders:
  Probing Higgsino Dark Matter}},  \href{https://arxiv.org/abs/2609.08712}{{\tt
  2609.08712}}.

\bibitem{Yuan:2026djt}
G.-W. Yuan, B.~Zhang, W.-Y. Cao, L.~Feng and R.~Yang, \emph{{ALP-mediated
  inelastic dark matter and the LUX-ZEPLIN high-recoil candidate event
  LZ230616}},  \href{https://arxiv.org/abs/2609.08893}{{\tt 2609.08893}}.

\bibitem{Zhu:2026dag}
P.~Zhu, G.~Dalla Valle~Garcia, X.-G. Wang, A.~W. Thomas and M.~J. White,
  \emph{{Endothermic dark matter with a light dark photon and the LUX--ZEPLIN
  high-energy nuclear-recoil candidate}},
  \href{https://arxiv.org/abs/2609.09015}{{\tt 2609.09015}}.

\bibitem{Asadi:2026iot}
P.~Asadi, A.~Batz, P.~J. Fox, S.~D. Homiller and G.~D. Kribs, \emph{{For Whom
  the Xenon Recoils: Magnetic Inelastic Dark Baryons}},
  \href{https://arxiv.org/abs/2609.09107}{{\tt 2609.09107}}.

\bibitem{Lee:2026xxh}
V.~S.~H. Lee and L.~Randall, \emph{{A Warped Extra Dimensional Candidate for
  the LZ 248 keV Event}},  \href{https://arxiv.org/abs/2609.09136}{{\tt
  2609.09136}}.

\bibitem{Lee:2026jxl}
S.~J. Lee and T.~Youn, \emph{{Mixing-suppressed inelastic dark matter: a
  minimal model for the LZ 248 keV event}},
  \href{https://arxiv.org/abs/2609.09138}{{\tt 2609.09138}}.

\bibitem{Langhoff:2026ujr}
K.~Langhoff, \emph{{Heavy Higgsino Interpretation of the LZ Event}},
  \href{https://arxiv.org/abs/2609.09385}{{\tt 2609.09385}}.

\bibitem{Chatterjee:2026scv}
A.~Chatterjee, D.~Das, S.~A. Pasha, A.~Pukhov and R.~Puri, \emph{{Radiative
  Corrections to the Direct Detection of Inelastic Scattering of Higgsino-like
  Neutralino Dark Matter}},  \href{https://arxiv.org/abs/2609.09830}{{\tt
  2609.09830}}.

\bibitem{Fan:2026hzw}
Z.-T. Fan, H.-J. He, Y.-C. Wang and Y.~Zhao, \emph{{Inelastic Dark Matter and
  High-Energy Recoil Signatures in LZ}},
  \href{https://arxiv.org/abs/2609.10491}{{\tt 2609.10491}}.

\bibitem{Qi:2026vyp}
X.~Qi and H.~Sun, \emph{{Nonthermal Solar Stalling of an Inelastic Scalar
  Signal in Xenon}},  \href{https://arxiv.org/abs/2609.10636}{{\tt
  2609.10636}}.

\bibitem{Frolovsky:2026tvq}
D.~Frolovsky and S.~V. Ketov, \emph{{Higgsino dark matter in the Starobinsky
  supergravity with the MSSM in light of the LUX-ZEPLIN event}},
  \href{https://arxiv.org/abs/2609.11241}{{\tt 2609.11241}}.

\bibitem{Kumar:2026lgi}
R.~Kumar and H.~K. Prajapati, \emph{{Generalized Chiral $U(1)_{B-L}$ with
  Inelastic Scalar Dark Matter for the LZ 248 keV Event}},
  \href{https://arxiv.org/abs/2609.10827}{{\tt 2609.10827}}.

\bibitem{Nguyen:2026lui}
T.~T.~Q. Nguyen, T.~Linden and D.~Hooper, \emph{{Solar Neutrino Constraints on
  Inelastic Dark Matter Scattering in Light of Recent LUX-ZEPLIN
  Observations}},  \href{https://arxiv.org/abs/2609.11833}{{\tt 2609.11833}}.

\bibitem{Sannino:2026hkc}
F.~Sannino and J.~Turner, \emph{{Interpreting the LZ 248 keV Event using Dark
  QCD}},  \href{https://arxiv.org/abs/2609.24988}{{\tt 2609.24988}}.

\bibitem{Maier:2026qpr}
B.~Maier and M.~Spannowsky, \emph{{The LHC is not enough: the LZ High-Recoil
  Event at FCC-hh and a Muon Collider}},
  \href{https://arxiv.org/abs/2609.26870}{{\tt 2609.26870}}.

\bibitem{Girmohanta:2026qzo}
S.~Girmohanta and T.~H. Jung, \emph{{Environment-dependent mass splitting to
  suppress solar capture in the inelastic-doublet interpretation of the LZ
  event}},  \href{https://arxiv.org/abs/2609.28129}{{\tt 2609.28129}}.

\bibitem{Tang:2026lgu}
J.~Tang and P.-H. Gu, \emph{{Inelastic dark matter mediated by natural dark
  photon}},  \href{https://arxiv.org/abs/2609.28415}{{\tt 2609.28415}}.

\bibitem{Nomura:2026qyq}
Y.~Nomura, \emph{{Dark Matter as the Z{\_}2 Partner of the Standard Model Higgs
  Boson}},  \href{https://arxiv.org/abs/2609.02505}{{\tt 2609.02505}}.

\bibitem{Ahmed:2026kan}
W.~Ahmed, A.~Ahmad and M.~U. Rehman, \emph{{Xenon Isotope Filtering at the
  Kinematic Edge of Inelastic Dark Matter}},
  \href{https://arxiv.org/abs/2609.15634}{{\tt 2609.15634}}.

\bibitem{Gemmell:2026yaw}
C.~Gemmell, D.~Hooper and G.~Krnjaic, \emph{{A Simple Dark Matter Model to
  Explain the LZ Event and Galactic Center Excess}},
  \href{https://arxiv.org/abs/2609.26570}{{\tt 2609.26570}}.

\bibitem{Delepine:2026ith}
D.~Delepine and S.~Khalil, \emph{{Model-Independent Sideband Constraints on
  Inelastic Dark Matter at the LZ High-Recoil Candidate}},
  \href{https://arxiv.org/abs/2609.26698}{{\tt 2609.26698}}.

\bibitem{Langhoff:2026qgo}
K.~Langhoff and H.~Xiao, \emph{{Sneaky Sneutrino Scattering at LZ}},
  \href{https://arxiv.org/abs/2609.28616}{{\tt 2609.28616}}.

\bibitem{Murayama:2026apt}
H.~Murayama and B.~Noether, \emph{{Heavy Inert Doublet Reconciles LZ and
  IceCube}},  \href{https://arxiv.org/abs/2609.28819}{{\tt 2609.28819}}.

\bibitem{Das:2026jsq}
A.~Das, S.~Majumdar, M.~Sen and A.~Srivastav, \emph{{GAIA meets LZ: a high
  velocity tail-tale sign for dark matter}},
  \href{https://arxiv.org/abs/2609.35737}{{\tt 2609.35737}}.

\bibitem{Roy:2026wzb}
S.~Roy, P.~Schwaller and C.~E.~M. Wagner, \emph{{Interpreting the High-Recoil
  LUX-ZEPLIN Event with Bino-/Singlino-like and Higgsino Dark Matter}},
  \href{https://arxiv.org/abs/2609.40234}{{\tt 2609.40234}}.

\bibitem{Unwin:2026rdp}
J.~Unwin, \emph{{Axion Portal Dark Matter and the LUX-ZEPLIN High-Recoil
  Event}},  \href{https://arxiv.org/abs/2609.04186}{{\tt 2609.04186}}.

\bibitem{Lou:2026idn}
Y.~Lou and C.-T. Lu, \emph{{Fermionic Dark Matter Absorption and the
  High-Energy Event in LUX-ZEPLIN}},
  \href{https://arxiv.org/abs/2609.01592}{{\tt 2609.01592}}.

\bibitem{Su:2026rwz}
L.~Su, J.~M. Yang and W.-N. Yang, \emph{{Inelastic Dark Matter Signature at
  High Recoil Energy in LUX-ZEPLIN and CRESST}},
  \href{https://arxiv.org/abs/2609.01475}{{\tt 2609.01475}}.

\bibitem{Yamashita:2026ump}
K.~Yamashita, \emph{{Inelastic Dark Photon Dark Matter for the LUX-ZEPLIN
  High-Recoil Event and the Galactic Halo Gamma-Ray Excess}},
  \href{https://arxiv.org/abs/2609.02868}{{\tt 2609.02868}}.

\bibitem{Elahi:2026vlm}
F.~Elahi and P.~Schwaller, \emph{{A Vector-Like Lepton Interpretation of the
  High-Energy Nuclear Recoil Candidate in LUX-ZEPLIN}},
  \href{https://arxiv.org/abs/2609.08993}{{\tt 2609.08993}}.

\bibitem{Khan:2026nwp}
I.~Khan, S.~Capozziello, G.~Mustafa, F.~Atamurotov, A.~Abdujabbarov and
  C.~Yuan, \emph{{Nuclear interference versus dark sector excitation in the 248
  keV LUX-ZEPLIN recoil candidate}},
  \href{https://arxiv.org/abs/2609.09230}{{\tt 2609.09230}}.

\bibitem{He:2026hqz}
Y.~He, \emph{{Transition magnetic-dipole dark matter and the LZ230616
  high-recoil candidate}},  \href{https://arxiv.org/abs/2609.10453}{{\tt
  2609.10453}}.

\bibitem{Heikinheimo:2026kwp}
M.~Heikinheimo and N.~Zimmermann, \emph{{Cosmic ray boosted dark matter with
  momentum dependent interactions can explain the LZ 248 keV event}},
  \href{https://arxiv.org/abs/2609.11600}{{\tt 2609.11600}}.

\bibitem{Chauhan:2026udz}
B.~Chauhan, S.~Sahasrabuddhe and M.~Sen, \emph{{A closer look at the LZ 248 keV
  event through the lens of cosmic-ray boosted dark matter}},
  \href{https://arxiv.org/abs/2609.24982}{{\tt 2609.24982}}.

\bibitem{Jung:2026otm}
H.~Jung and S.~C. Park, \emph{{Dark Diffraction at LZ from a Screened Neutral
  Composite Baryon}},  \href{https://arxiv.org/abs/2609.25723}{{\tt
  2609.25723}}.

\bibitem{Khan:2026osp}
I.~Khan, S.~Capozziello, G.~Mustafa, F.~Botirov, A.~Abdujabbarov, F.~Atamurotov
  et~al., \emph{{Elastic toroidal vector dark matter through a dark photon in
  the LUX-ZEPLIN high recoil window}},
  \href{https://arxiv.org/abs/2609.25114}{{\tt 2609.25114}}.

\bibitem{Capolupo:2026uyi}
A.~Capolupo, G.~Pisacane and R.~Serao, \emph{{Solar capture and annihilation of
  fermionic and scalar dark matter motivated by the LUX-ZEPLIN high-recoil
  candidate}},  \href{https://arxiv.org/abs/2609.28218}{{\tt 2609.28218}}.

\bibitem{Liang:2026coz}
J.-H. Liang, Z.~Liu, V.~Q. Tran and Y.~Xu, \emph{{LZ Nuclear-Recoil Excess from
  Boosted Light Magnetic Dipole-dipole Dark Matter}},
  \href{https://arxiv.org/abs/2609.06756}{{\tt 2609.06756}}.

\bibitem{Okada:2026eol}
N.~Okada and O.~Seto, \emph{{Inelastic $B-L$ scalar dark matter and the
  LUX-ZEPLIN event}},  \href{https://arxiv.org/abs/2609.06909}{{\tt
  2609.06909}}.

\bibitem{Kannike:2026qyl}
K.~Kannike, M.~Raidal and A.~Strumia, \emph{{Boosted dark particles and the LZ
  nuclear recoil event}},  \href{https://arxiv.org/abs/2609.07742}{{\tt
  2609.07742}}.

\bibitem{Chattaraj:2026fxn}
A.~Chattaraj, A.~Majumdar, D.~K. Papoulias and R.~Srivastava, \emph{{Can
  Elastic Neutrino Scattering Account for the LZ230616 Event?}},
  \href{https://arxiv.org/abs/2609.10504}{{\tt 2609.10504}}.

\bibitem{Aghaie:2026vsu}
M.~Aghaie and A.~Strumia, \emph{{Neutron disappearance and the LZ nuclear
  recoil event}},  \href{https://arxiv.org/abs/2609.09037}{{\tt 2609.09037}}.

\bibitem{Brdar:2026ukx}
V.~Brdar and D.~S. Chattopadhyay, \emph{{A Narrow Neutrino Window for the LZ
  Event}},  \href{https://arxiv.org/abs/2609.30255}{{\tt 2609.30255}}.

\bibitem{Kalita:2026hvb}
J.~Kalita, \emph{{Limits on Primordial Black Hole Evaporation from
  LUX-ZEPLIN}},  \href{https://arxiv.org/abs/2609.29462}{{\tt 2609.29462}}.

\bibitem{Chanda:2026wdl}
P.~Chanda, B.~Das and S.~Mukherjee, \emph{{Does the High-Energy LUX--ZEPLIN
  Event Suggest Hyperfine Atomic Dark Matter?}},
  \href{https://arxiv.org/abs/2609.36814}{{\tt 2609.36814}}.

\bibitem{Bishara:2016hek}
F.~Bishara, J.~Brod, B.~Grinstein and J.~Zupan, \emph{{Chiral Effective Theory
  of Dark Matter Direct Detection}},
  \href{http://dx.doi.org/10.1088/1475-7516/2017/02/009}{\emph{JCAP} {\bf 02}
  (2017) 009}, [\href{https://arxiv.org/abs/1611.00368}{{\tt 1611.00368}}].

\bibitem{Brod:2017bsw}
J.~Brod, A.~Gootjes-Dreesbach, M.~Tammaro and J.~Zupan, \emph{{Effective Field
  Theory for Dark Matter Direct Detection up to Dimension Seven}},
  \href{http://dx.doi.org/10.1007/JHEP10(2018)065}{\emph{JHEP} {\bf 10} (2018)
  065}, [\href{https://arxiv.org/abs/1710.10218}{{\tt 1710.10218}}].

\bibitem{Bhattacharya:2021edh}
S.~Bhattacharya and J.~Wudka, \emph{{Effective theories with dark matter
  applications}}, \href{http://dx.doi.org/10.1142/S0218271821300044}{\emph{Int.
  J. Mod. Phys. D} {\bf 30} (2021) 2130004},
  [\href{https://arxiv.org/abs/2104.01788}{{\tt 2104.01788}}].

\bibitem{Liang:2023yta}
J.-H. Liang, Y.~Liao, X.-D. Ma and H.-L. Wang, \emph{{Dark sector effective
  field theory}}, \href{http://dx.doi.org/10.1007/JHEP12(2023)172}{\emph{JHEP}
  {\bf 12} (2023) 172}, [\href{https://arxiv.org/abs/2309.12166}{{\tt
  2309.12166}}].

\bibitem{ATLAS:2021kxv}
{\scshape ATLAS} collaboration, G.~Aad et~al., \emph{{Search for new phenomena
  in events with an energetic jet and missing transverse momentum in $pp$
  collisions at $\sqrt {s}$ =13 TeV with the ATLAS detector}},
  \href{http://dx.doi.org/10.1103/PhysRevD.103.112006}{\emph{Phys. Rev. D} {\bf
  103} (2021) 112006}, [\href{https://arxiv.org/abs/2102.10874}{{\tt
  2102.10874}}].

\bibitem{Racco_2015}
D.~Racco, A.~Wulzer and F.~Zwirner, \emph{Robust collider limits on
  heavy-mediator dark matter},
  \href{http://dx.doi.org/10.1007/jhep05(2015)009}{\emph{Journal of High Energy
  Physics} {\bf 2015} (May, 2015) }.

\bibitem{De_Simone_2016}
A.~De~Simone and T.~Jacques, \emph{Simplified models vs. effective field theory
  approaches in dark matter searches},
  \href{http://dx.doi.org/10.1140/epjc/s10052-016-4208-4}{\emph{The European
  Physical Journal C} {\bf 76} (2016) }.

\bibitem{Bauer:2016pug}
M.~Bauer, A.~Butter, N.~Desai, J.~Gonzalez-Fraile and T.~Plehn, \emph{{Validity
  of dark matter effective theory}},
  \href{http://dx.doi.org/10.1103/PhysRevD.95.075036}{\emph{Phys. Rev. D} {\bf
  95} (2017) 075036}, [\href{https://arxiv.org/abs/1611.09908}{{\tt
  1611.09908}}].

\bibitem{Bruggisser:2016nzw}
S.~Bruggisser, F.~Riva and A.~Urbano, \emph{{The Last Gasp of Dark Matter
  Effective Theory}},
  \href{http://dx.doi.org/10.1007/JHEP11(2016)069}{\emph{JHEP} {\bf 11} (2016)
  069}, [\href{https://arxiv.org/abs/1607.02475}{{\tt 1607.02475}}].

\bibitem{Pobbe:2017wrj}
F.~Pobbe, A.~Wulzer and M.~Zanetti, \emph{{Setting limits on Effective Field
  Theories: the case of Dark Matter}},
  \href{http://dx.doi.org/10.1007/JHEP08(2017)074}{\emph{JHEP} {\bf 08} (2017)
  074}, [\href{https://arxiv.org/abs/1704.00736}{{\tt 1704.00736}}].

\bibitem{Belwal:2017nkw}
S.~Belwal, M.~Drees and J.~S. Kim, \emph{{Analysis of the Bounds on Dark Matter
  Models from Monojet Searches at the LHC}},
  \href{http://dx.doi.org/10.1103/PhysRevD.98.055017}{\emph{Phys. Rev. D} {\bf
  98} (2018) 055017}, [\href{https://arxiv.org/abs/1709.08545}{{\tt
  1709.08545}}].

\bibitem{Belyaev:2018pqr}
A.~Belyaev, E.~Bertuzzo, C.~Caniu~Barros, O.~Eboli, G.~Grilli Di~Cortona,
  F.~Iocco et~al., \emph{{Interplay of the LHC and non-LHC Dark Matter searches
  in the Effective Field Theory approach}},
  \href{http://dx.doi.org/10.1103/PhysRevD.99.015006}{\emph{Phys. Rev. D} {\bf
  99} (2019) 015006}, [\href{https://arxiv.org/abs/1807.03817}{{\tt
  1807.03817}}].

\bibitem{Roy:2024ear}
A.~Roy, B.~Dasgupta and M.~Guchait, \emph{{Constraining Asymmetric Dark Matter
  using colliders and direct detection}},
  \href{http://dx.doi.org/10.1007/JHEP08(2024)095}{\emph{JHEP} {\bf 08} (2024)
  095}, [\href{https://arxiv.org/abs/2402.17265}{{\tt 2402.17265}}].

\bibitem{Roy:2025pht}
A.~Roy, M.~A. Schmidt and G.~Valencia, \emph{{Monojet and direct detection
  constraints on real scalar dark matter: EFT and a simple UV completion}},
  \href{http://dx.doi.org/10.1007/JHEP02(2026)167}{\emph{JHEP} {\bf 02} (2026)
  167}, [\href{https://arxiv.org/abs/2509.14869}{{\tt 2509.14869}}].

\bibitem{CMS:2021far}
{\scshape CMS} collaboration, A.~Tumasyan et~al., \emph{{Search for new
  particles in events with energetic jets and large missing transverse momentum
  in proton-proton collisions at $ \sqrt{s} $ = 13 TeV}},
  \href{http://dx.doi.org/10.1007/JHEP11(2021)153}{\emph{JHEP} {\bf 11} (2021)
  153}, [\href{https://arxiv.org/abs/2107.13021}{{\tt 2107.13021}}].

\bibitem{Anand:2013yka}
N.~Anand, A.~L. Fitzpatrick and W.~C. Haxton, \emph{{Weakly interacting massive
  particle-nucleus elastic scattering response}},
  \href{http://dx.doi.org/10.1103/PhysRevC.89.065501}{\emph{Phys. Rev. C} {\bf
  89} (2014) 065501}, [\href{https://arxiv.org/abs/1308.6288}{{\tt
  1308.6288}}].

\bibitem{Bose:2026szs}
D.~Bose, A.~K. Saha, N.~Raj, T.~N. Maity and R.~Laha, \emph{{LUX-ZEPLIN's
  Stairway to Hea$\nu$en: limits on elastic scatters of dark matter from solar
  capture}},  \href{https://arxiv.org/abs/2609.21823}{{\tt 2609.21823}}.

\bibitem{Fitzpatrick:2012ix}
A.~L. Fitzpatrick, W.~Haxton, E.~Katz, N.~Lubbers and Y.~Xu, \emph{{The
  Effective Field Theory of Dark Matter Direct Detection}},
  \href{http://dx.doi.org/10.1088/1475-7516/2013/02/004}{\emph{JCAP} {\bf 02}
  (2013) 004}, [\href{https://arxiv.org/abs/1203.3542}{{\tt 1203.3542}}].

\bibitem{Gorton:2022eed}
O.~C. Gorton, C.~W. Johnson, C.~Jiao and J.~Nikoleyczik, \emph{{dmscatter: A
  fast program for WIMP-nucleus scattering}},
  \href{http://dx.doi.org/10.1016/j.cpc.2022.108597}{\emph{Comput. Phys.
  Commun.} {\bf 284} (2023) 108597},
  [\href{https://arxiv.org/abs/2209.09187}{{\tt 2209.09187}}].

\bibitem{deSalas:2019pee}
P.~F. de~Salas, K.~Malhan, K.~Freese, K.~Hattori and M.~Valluri, \emph{{On the
  estimation of the Local Dark Matter Density using the rotation curve of the
  Milky Way}},
  \href{http://dx.doi.org/10.1088/1475-7516/2019/10/037}{\emph{JCAP} {\bf 10}
  (2019) 037}, [\href{https://arxiv.org/abs/1906.06133}{{\tt 1906.06133}}].

\bibitem{Baxter:2021pqo}
D.~Baxter et~al., \emph{{Recommended conventions for reporting results from
  direct dark matter searches}},
  \href{http://dx.doi.org/10.1140/epjc/s10052-021-09655-y}{\emph{Eur. Phys. J.
  C} {\bf 81} (2021) 907}, [\href{https://arxiv.org/abs/2105.00599}{{\tt
  2105.00599}}].

\bibitem{Alloul:2013bka}
A.~Alloul, N.~D. Christensen, C.~Degrande, C.~Duhr and B.~Fuks,
  \emph{{FeynRules 2.0 - A complete toolbox for tree-level phenomenology}},
  \href{http://dx.doi.org/10.1016/j.cpc.2014.04.012}{\emph{Comput. Phys.
  Commun.} {\bf 185} (2014) 2250--2300},
  [\href{https://arxiv.org/abs/1310.1921}{{\tt 1310.1921}}].

\bibitem{Cirelli:2013ufw}
M.~Cirelli, E.~Del~Nobile and P.~Panci, \emph{{Tools for model-independent
  bounds in direct dark matter searches}},
  \href{http://dx.doi.org/10.1088/1475-7516/2013/10/019}{\emph{JCAP} {\bf 10}
  (2013) 019}, [\href{https://arxiv.org/abs/1307.5955}{{\tt 1307.5955}}].

\bibitem{DelNobile:2021wmp}
E.~Del~Nobile, \emph{{The Theory of Direct Dark Matter Detection: A Guide to
  Computations}},  \href{https://arxiv.org/abs/2104.12785}{{\tt 2104.12785}}.

\bibitem{Roy:2025inq}
A.~Roy and R.~R. Volkas, \emph{{Muonphilic asymmetric dark matter at a future
  muon collider}}, \href{http://dx.doi.org/10.1007/JHEP04(2026)157}{\emph{JHEP}
  {\bf 04} (2026) 157}, [\href{https://arxiv.org/abs/2512.23159}{{\tt
  2512.23159}}].

\bibitem{Guchait:2020wqn}
M.~Guchait and A.~Roy, \emph{{Light Singlino Dark Matter at the LHC}},
  \href{http://dx.doi.org/10.1103/PhysRevD.102.075023}{\emph{Phys. Rev. D} {\bf
  102} (2020) 075023}, [\href{https://arxiv.org/abs/2005.05190}{{\tt
  2005.05190}}].

\bibitem{Busoni:2013lha}
G.~Busoni, A.~De~Simone, E.~Morgante and A.~Riotto, \emph{{On the Validity of
  the Effective Field Theory for Dark Matter Searches at the LHC}},
  \href{http://dx.doi.org/10.1016/j.physletb.2013.11.069}{\emph{Phys. Lett. B}
  {\bf 728} (2014) 412--421}, [\href{https://arxiv.org/abs/1307.2253}{{\tt
  1307.2253}}].

\bibitem{Alwall:2014hca}
J.~Alwall, R.~Frederix, S.~Frixione, V.~Hirschi, F.~Maltoni, O.~Mattelaer
  et~al., \emph{{The automated computation of tree-level and next-to-leading
  order differential cross sections, and their matching to parton shower
  simulations}}, \href{http://dx.doi.org/10.1007/JHEP07(2014)079}{\emph{JHEP}
  {\bf 07} (2014) 079}, [\href{https://arxiv.org/abs/1405.0301}{{\tt
  1405.0301}}].

\bibitem{Sjostrand:2014zea}
T.~Sj\"ostrand, S.~Ask, J.~R. Christiansen, R.~Corke, N.~Desai, P.~Ilten
  et~al., \emph{{An introduction to PYTHIA 8.2}},
  \href{http://dx.doi.org/10.1016/j.cpc.2015.01.024}{\emph{Comput. Phys.
  Commun.} {\bf 191} (2015) 159--177},
  [\href{https://arxiv.org/abs/1410.3012}{{\tt 1410.3012}}].

\bibitem{deFavereau:2013fsa}
{\scshape DELPHES 3} collaboration, J.~de~Favereau, C.~Delaere, P.~Demin,
  A.~Giammanco, V.~Lema\^\i{}tre, A.~Mertens et~al., \emph{{DELPHES 3, A
  modular framework for fast simulation of a generic collider experiment}},
  \href{http://dx.doi.org/10.1007/JHEP02(2014)057}{\emph{JHEP} {\bf 02} (2014)
  057}, [\href{https://arxiv.org/abs/1307.6346}{{\tt 1307.6346}}].

\end{thebibliography}\endgroup

\end{document}